\documentclass[aps,a4paper,amssymb,amsmath,reprint,showpacs,superscriptaddress]{revtex4-1}

\usepackage{amsmath,amssymb}
\usepackage[hiresbb,pdftex]{graphicx}
\usepackage[pdftex]{xcolor}
\usepackage{bm,braket,latexsym,multirow}
\usepackage{cancel}

\usepackage[pdftex,
draft=false,
bookmarks=true,
bookmarksnumbered=false,
bookmarksopen=false,
bookmarksopenlevel=4,
colorlinks=true,
anchorcolor=black,
citecolor=blue,
filecolor=magenta,
linkcolor=blue,
linkbordercolor={1 0 0},
urlcolor=violet,
pdfborder={0 0 1},
pdftitle={},
pdfauthor={},
pdfsubject={},
pdfkeywords={},
pdfpagemode=UseThumbs,
]{hyperref}

\newcommand{\bma}[0]{BaMn$_2$As$_2$}
\newcommand{\Pa}[0]{$\mathcal{P}$}
\newcommand{\T}[0]{$\mathcal{T}$}
\newcommand{\PT}[0]{$\mathcal{PT}$}
\newcommand{\static}[0]{\xrightarrow{\text{sta.}}}
\newcommand{\mr}[2]{#1 \, {\mathrm{ \,  #2 \,}}}
\newcommand{\trace}[1]{\mathrm{Tr}[{#1}]}
\newcommand{\impave}[1]{\Braket{#1}_\text{imp}}
\newcommand{\bk}{\bm{k}}

\begin{document}

\title{Nonlinear electric transport in odd-parity magnetic multipole systems: \\ Application to Mn-based compounds}
\author{Hikaru Watanabe}
\email[]{watanabe.hikaru.43n@st.kyoto-u.ac.jp}
\affiliation{Department of Physics, Graduate School of Science, Kyoto University, Kyoto 606-8502, Japan}
\author{Youichi Yanase}
\affiliation{Department of Physics, Graduate School of Science, Kyoto University, Kyoto 606-8502, Japan}
\affiliation{Institute for Molecular Science, Okazaki,444-8585, Japan}
\date{\today}

\begin{abstract}
Violation of parity symmetry gives rise to various physical phenomena such as nonlinear transport and cross-correlated responses. In particular, the nonlinear conductivity has been attracting a lot of attentions  in spin-orbit coupled semiconductors, superconductors, topological materials, and so on. In this paper we present theoretical study of the nonlinear conductivity in odd-parity magnetic multipole ordered systems whose \PT{}-symmetry is essentially distinct from the previously studied acentric systems. Combining microscopic formulation and symmetry analysis, we classify the nonlinear responses in the \PT{}-symmetric systems as well as \T{}-symmetric (non-magnetic) systems, and uncover nonlinear conductivity unique to the odd-parity magnetic multipole systems. A giant nonlinear Hall effect, nematicity-assisted dichroism and magnetically-induced Berry curvature dipole effect are proposed and demonstrated in a model for Mn-based magnets.   

\end{abstract}
\maketitle

\section{Introduction}\label{Sec_introduction}

Nonlinear responses have been giving rise to a lot of research interest in condensed matter physics. For instance, the nonlinear optical response provides a powerful tool for spectrometry. It has been used to obtain a real-space imaging of the parity-violating magnetic order in insulators~\cite{Fiebig2005SHG_review,VanAken2007ToroidalDomain} and to explore exotic order in spin-orbit coupled metals and superconductors~\cite{Petersen2006SHG,Zhao2015SHG_Sr2IrO4,zhao2017global,Harter2017SHG_Cd2Re2O7}. In the optical nonlinear responses, energy of irradiating light  is usually larger than that of electron bands, and the observed signals are attributed to interband transitions~\cite{Sturman1992Book}. On the other hand, intraband transitions are also important in conductivity measurements in which the frequency of the probe is usually lower than that of optical probes and comparable to the electronic energy scale. It is therefore expected that nonlinear responses are informative for investigating metallic compounds where the intraband transitions are relevant. 

Regarding the nonlinear response in metals, the second-order nonlinear conductivity (NLC) measurement has attracted much attention. Previous studies are mainly divided into two streams; field-induced NLC and field-free NLC. The former can be traced back to Rikken's seminal works~\cite{Rikken2001magnetochiral_anisotropy,Rikken2005magnetoelectric_anisotropy}. They realized the longitudinal NLC under an external magnetic field, and significant enhancement has recently been discovered in strongly spin-orbit coupled semiconductors and superconductors~\cite{Tokura2018nonreciprocal_review}. The microscopic origin of the longitudinal NLC is attributed to a semiclassical contribution which we call Drude term~\cite{Ideue2017,Wakatsuki2017,Itahashi2020STOnonreciprocal}. On the other hand, a lot of theoretical and experimental efforts have recently been devoted to the transverse NLC, that is, nonlinear Hall effect~\cite{Sodemann2015,Xu2018BCD_switchable,Ma2019BCD_experiment_WTe2}. The nonlinear Hall effect realized without the magnetic field is rooted in a geometric quantity named Berry curvature dipole (BCD)~\cite{Sodemann2015}.

According to the symmetry argument, the second-order NLC requires violation of parity symmetry \Pa{}. The condition is satisfied by the acentric property of crystals which were previously studied~\cite{Rikken2001magnetochiral_anisotropy,Rikken2005magnetoelectric_anisotropy,Tokura2018nonreciprocal_review,Ideue2017,Sodemann2015,Xu2018BCD_switchable,Ma2019BCD_experiment_WTe2}. In contrast, the \Pa{}-symmetry breaking can also be accompanied by the magnetic order, that is called odd-parity magnetic multipole order~\cite{Spaldin2008a,Watanabe2018grouptheoretical,Hayami2018Classification}. It is expected that counterparts of the NLC exist in magnetic metals.

A key to the odd-parity magnetic multipole order is locally-noncentrosymmetric property of crystals. With such structure of crystals, the local site-symmetry of atoms does not have \Pa{}-symmetry although the global \Pa{}-symmetry is preserved owing to the sublattice degree of freedom~\cite{Maruyama2012,Yanase2014zigzag}. Then, the anti-symmetric spin-orbit coupling  (ASOC) emerges in a sublattice dependent way and gives rise to exotic responses such as the antiferromagnetic Edelstein effect~\cite{Yanase2014zigzag,Zelezny2014NeelorbitTorque,Hayami2014h,hikaruwatanabe2017}. Supposing the antiferromagnetic order preserving the translational symmetry, both of the \Pa{} and \T{}-symmetries may be violated while the combined symmetry, namely \PT{}-symmetry, is preserved. Such parity-violating but \PT{}-symmetric magnetic order is called odd-parity magnetic multipole order and has been discussed in the context of multipole physics~\cite{Watanabe2018grouptheoretical,Hayami2018Classification} and antiferromagnetic spintronics~\cite{Jungwirth2016,Manchon2019spin-orbit-torque_review,Watanabe2018}. More than 100 candidate materials such as BaMn$_2$As$_2$ and EuMnBi$_2$ have been identified~\cite{Watanabe2018grouptheoretical}.

In this work, the NLC in odd-parity magnetic multipole metals are investigated. We present a general symmetry classification of NLC based on a quantum mechanical calculation. Supported by the microscopic analysis, we clarify NLC characteristic of magnetic metals with and without external magnetic field. We find that the NLC at $H=0$ is a measure of the ASOC. Furthermore, we reveal two types of field-induced NLC; the nematicity-assisted dichroism, and Berry curvature dipole effect induced by what we call magnetic ASOC. These phenomena originate from locally-noncentrosymmetric crystal structures and magnetic order, and hence have striking difference from the NLC in noncentrosymmetric (non-magnetic) crystals. We show the correspondence between \T{}-symmetric and \PT{}-symmetric systems in Table~\ref{Table_NLC_MagneticField}.  

		\begin{table}[htbp]
		\caption{Second-order NLC in \T{}/\PT{}-symmetric systems with/without magnetic field $\bm{H}$. Dominant contributions such as the BCD term are shown. The boldfaced terms are studied in this work.}
		\label{Table_NLC_MagneticField}
		\centering
		\begin{tabular}{l|cc}
					&\T{}	&\PT{}	 \\ \hline
		$\bm{H}=0$	&BCD			&\textbf{Drude}	 \\
		$\bm{H}\ne 0$	&magnetic Drude	&\textbf{magnetic BCD}	 
		\end{tabular}
		\end{table}

\section{Quantum theory and symmetry analysis}\label{Sec_quantumtheory_symmetryanalysis}

A theoretical treatment of the second-order NLC has been established in Sipe and his coworkers' works~\cite{Sipe1993,Aversa1995,Sipe2000secondorder} where the nonlinear response functions are derived from straightforward extension of the linear response theory~\cite{Kubo1957}. A detailed calculation of the NLC is shown in Appendix~\ref{App_Sec_derivation_NLC}. Here we only describe the outline of derivation.

A spatially-uniform electric field $\bm{E}(t)$ is introduced so as to be compatible with calculations based on the Bloch states in the length gauge
framework~\cite{Sipe1993,Aversa1995,Sipe2000secondorder,Ventura2017,Passos2018,Parker2019}. Using the density matrix formalism, we derive the second-order electric current in the frequency domain as 
		\begin{align}
		J^{\mu (2)} (\omega) = \sum_{\nu\lambda} &\int \frac{d\omega_1d\omega_2}{2\pi}  \sigma^{\mu;\nu\lambda} (\omega; \omega_1,\omega_2 ) \notag \\
			&\times E^\nu \left( \omega_1  \right)E^\lambda \left( \omega_2  \right) \delta (\omega-\omega_1-\omega_2).
		\end{align}
Assuming the clean limit where the relaxation time $\tau \rightarrow \infty$ within the transport regime $\omega\tau \ll 1$, the NLC is classified by the dependence on the phenomenological relaxation time $\tau = \gamma^{-1}$ as
		\begin{equation}
		\sigma^{\mu;\nu \lambda} = \sigma^{\mu;\nu \lambda}_\text{D} + \sigma^{\mu;\nu \lambda}_\text{BCD} + \sigma^{\mu;\nu \lambda}_\text{int},
		\label{nonlinear_conductivity}
		\end{equation}
in which the indices $\nu,~\lambda$ of applied electric fields are symmetric. The first two components are obtained as
		\begin{align}
		\sigma^{\mu;\nu \lambda}_\text{D} 
				&= - \frac{e^3}{\gamma^2}\int \frac{d\bm{k}}{\left( 2\pi \right)^d} \sum_a \partial_\mu \partial_\nu \partial_\lambda  \epsilon_{\bm{k}a}   f(\epsilon_{\bm{k}a}), \label{drude}\\
		\sigma^{\mu;\nu \lambda}_\text{BCD} 
				&= \frac{e^3}{2\gamma}\int \frac{d\bm{k}}{\left( 2\pi \right)^d} \sum_{a} \epsilon_{\mu\nu\kappa}   f(\epsilon_{\bm{k}a}) \partial_\lambda \Omega^\kappa_a + \left[ \nu \leftrightarrow \lambda \right],\\
				&= \frac{e^3}{2\gamma} \epsilon_{\mu\nu\kappa}   \mathcal{D}^{\,\lambda\kappa} + \left[ \nu \leftrightarrow \lambda \right], \label{BCD_term}
		\end{align}
which are the Drude~\cite{Ideue2017} and BCD~\cite{Sodemann2015,deyo2009semiclassical,Moore2010} terms, respectively. The index $a$ represents the band index. We introduced the electron charge $e>0$ and the BCD defined as
		\begin{equation}
		\mathcal{D}^{\,\mu\nu} = \int \frac{d\bm{k}}{\left( 2\pi \right)^d} \sum_{a} f(\epsilon_{\bm{k}a}) \partial_\mu \Omega^\nu_a.
		\end{equation}
These two terms are finite only in the metal state and divergent in the clean limit since both are Fermi surface terms. The remaining term $\sigma^{\mu;\nu \lambda}_\text{int}$ comes from interband transitions, and it is not divergent in the clean limit~\cite{Gao2014}. This term, therefore, gives a negligible contribution to the NLC in good metals.

In the group-theoretical classification of quantum phases, the parity-violating phases are classified into odd-parity electric/magnetic multipole phases where \T{}/\PT{}-symmetry is preserved~\cite{Watanabe2018grouptheoretical,Hayami2018Classification}. It is known that these preserved symmetries impose strong constraints on the response functions~\cite{hikaruwatanabe2017,Zelezny2017} in addition to equilibrium properties of the systems~\cite{cracknell2016magnetism}. Thus, the symmetry analysis enables us to classify the NLC allowed in either \T{}-symmetric or \PT{}-symmetric systems based on the relaxation time dependence. The result is shown in Table~\ref{Table_relaxation_time_dependence_2nd_conductivity}.

In \T{}-symmetric systems, all the terms in NLC are scaled by odd-order $O (\tau^{2n+1})$. Recalling the linear response theory, the scattering rate $\gamma$ can be replaced by the adiabaticity parameter whose sign represents irreversibility due to external fields~\cite{Kubo1957}. Thus, the NLC should be accompanied by a dissipative response. This is consistent with previous theories~\cite{Morimoto2018NonreciprocalElectronCorrelation,Hamamoto2019RachetScaling}. In contrast to the familiar linear conductivity, the Drude term is prohibited because it is even-order with respect to $\tau$. The leading order term is the BCD term for the transverse NLC.

 On the other hand, the \T{}-symmetry is broken by the magnetic order in the parity-violating \PT{}-symmetric systems which we focus on. Therefore, the relaxation time dependence is even-order $O(\tau^{2n})$, and intrinsic contributions $O(\tau^{0})$ are allowed. The leading order term is the Drude term $O(\tau^{2})$. We will show that the Drude term is a measure of the hidden ASOC characteristic of locally-noncentrosymmetric systems. The BCD term is prohibited to be consistent with the fact that the Berry curvature itself disappears due to the \PT{}-symmetry. Although the effect of the \T{}-symmetry breaking in acentric systems has been discussed in previous works~\cite{Gao2014,Sodemann2015,Nandy2019,Du2019}, our classification has clarified the contrasting role of \T{} and \PT{}-symmetries in NLC. Below we will see that the \PT{}-symmetry gives a clear insight into the NLC.

In our classification, extrinsic contributions such as the side jump and skew scattering are not taken into account~\cite{Nandy2019,Du2019,CXiao2019ModifiedSemiclassics,Du2020quantumTheoryofNHE}. We, however, note that the extrinsic contributions may be similarly classified by the symmetries. Indeed, for nonmagnetic impurities with $\delta$-function potential, we show that while extrinsic terms are allowed in the \T{}-symmetric systems~\cite{Du2019}, they are strongly suppressed by the \PT{}-symmetry (See Appendix~\ref{App_Sec_extrinsic}). This suppression is highly contrasting to the fact that the impurities play an important role in the NLC in \T{}-symmetric materials such as WTe$_2$~\cite{Kang2019}. When we focus on the \PT{}-symmetric magnetic systems, the classification in Table~\ref{Table_relaxation_time_dependence_2nd_conductivity} is meaningful beyond the relaxation time approximation for impurity scattering.
 
		\begin{table}[htbp]
		\caption{Relaxation time dependence of the second-order NLC in \T{}/\PT{}-symmetric systems. `N/A' denotes that the component is forbidden by symmetry.}
		\label{Table_relaxation_time_dependence_2nd_conductivity}
		\centering
		$
				\begin{array}{c|ccc}
				&\sigma_\text{D}	&\sigma_\text{BCD}	&\sigma_\text{int}	\\ \hline
				\text{\T{}}	&\text{N/A}&O(\tau)&O(\tau^{-1})\\
				\text{\PT{}}&O(\tau^2) &\text{N/A} &O(\tau^{0})
				\end{array}
		$
		\end{table}

All the terms in NLC are allowed in the absence of both \T{} and \PT{}-symmetry. For instance, the Drude term becomes finite when we apply magnetic fields to originally \T{}-symmetric systems~\cite{Rikken2001magnetochiral_anisotropy,Rikken2005magnetoelectric_anisotropy,Tokura2018nonreciprocal_review,Ideue2017}, that is described as `magnetic Drude' in Table~\ref{Table_NLC_MagneticField}. Similarly, we expect a magnetic-field-induced NLC in originally \PT{}-symmetric systems; the BCD term indeed arises from the \PT{}-symmetry breaking (called `magnetic BCD' in Table~\ref{Table_NLC_MagneticField}). This term is clarified in this work below. In the following, we consider the \PT{}-preserving antiferromagnetic metal with or without the magnetic field, and discuss the Drude and BCD terms which are dominant in clean metals.

\section{NLC in odd-parity magnetic multipole systems}\label{Sec_NLC_odd-parity_magnetic_multipole}

We introduce a minimal model of \bma{} which undergoes odd-parity magnetic multipole order~\cite{hikaruwatanabe2017}. Many magnetic compounds in the list of Ref.~\cite{Watanabe2018grouptheoretical} belong to the same class. The Hamiltonian reads
		\begin{equation}
			H(\bm{k}) =
					\epsilon(\bm{k}) \, \tau_0+ \bm{g} \left( \bm{k} \right) \cdot \bm{\sigma} \, \tau_z +\bm{h} \cdot \bm{\sigma} \, \tau_0 +  V_{\rm AB} (\bm{k}) \, \tau_x, \label{BMA_model_Hamiltonian}
		\end{equation}
where $\bm{\sigma}$ and $\bm{\tau}$ are Pauli matrices representing the spin and sublattice degrees of freedom, respectively. In addition to the intra-sublattice and inter-sublattice hoppings, $\epsilon (\bm{k})$ and $V_\text{AB}  (\bm{k})$, we introduce the staggered $g$-vector $\bm{g}  (\bm{k})= \bm{g}_0 (\bm{k}) + \bm{h}_\text{AF}$ consisting of the sublattice-dependent ASOC $\bm{g}_0  (\bm{k})$~\cite{Yanase2014zigzag,Zelezny2014NeelorbitTorque,Hayami2014h} and the molecular field $\bm{h}_\text{AF}= h_\text{AF} \hat{z}$ due to antiferromagnetic order in \bma{} ~\cite{Singh2009BaMn2As2_1,Singh2009BaMn2As2_2,Ramsal2013BMA_MagneticStructure}. The detailed material property of \bma{} and expressions of $\epsilon (\bm{k})$, $V_\text{AB}  (\bm{k})$, and $\bm{g}_0 (\bm{k})$ are available in Appendix~\ref{App_Sec_model_hamiltonian}. We also consider an external magnetic field $\bm{h}$ to discuss field-induced NLC. 

\subsection{Field-free nonlinear Hall effect}\label{Sec_nonlinear_Hall_no_field}

First, we show the NLC at zero magnetic field ($\bm{h} =\bm{0}$). Then, the NLC is mainly given by the Drude term (see Table~\ref{Table_relaxation_time_dependence_2nd_conductivity}), and it is determined by the anti-symmetric and anharmonic property of the energy dispersion [see Eq.~\eqref{drude}]. Such dispersion is known to be a pronounced property of the odd-parity magnetic multipole systems~\cite{Yanase2014zigzag,Hayami2014h,Sumita2016,hikaruwatanabe2017,Watanabe2018grouptheoretical}. In the case of \bma{}, the anti-symmetric component was identified to be a cubic term $k_xk_yk_z$~\cite{hikaruwatanabe2017}. Indeed, the energy spectrum of the model Eq.~\eqref{BMA_model_Hamiltonian} is obtained as
		\begin{equation}
		E^\pm_{\bm{k}}= \epsilon (\bm{k}) \pm \sqrt{V_{\rm AB}(\bm{k})^2 + \bm{g}(\bm{k})^2 }. \label{energyspectrum_no_magnetic_field}
		\end{equation}
The anti-symmetric distortion in the band structure arises from the coupling term $\bm{g}_0 (\bm{k}) \cdot \bm{h}_\text{AF}$ which is approximated by $\sim k_xk_yk_z$ near time-reversal-invariant momentum. Thus, $\sigma^{z;xy}$ and its cyclic components of NLC tensor are allowed. This indicates the nonlinear Hall effect, namely, the second-order electric current $J^z$ generated from the electric field $\bm{E} \parallel [110]$. For the strong antiferromagnet, $|\bm{h}_\text{AF}| \gg |\epsilon (\bm{k})|$, $|V_\text{AB}  (\bm{k})|$, $|\bm{g}_0 (\bm{k})|$, the Drude component is analytically obtained as 
 		\begin{equation}
		\sigma^{z;xy}_\text{D} =\sigma^{x;yz}_\text{D} =\sigma^{y;zx}_\text{D} = \frac{e^3\alpha_{\parallel}n }{4\gamma^2} \,\text{sgn\,} (h_\text{AF}), \label{Drude_no_external_field}
		\end{equation}
in the lightly-hole-doped region. Here $n$ denotes the carrier density of holes and $\alpha_{\parallel}$ represents the strength of ASOC parallel to the staggered magnetization $\bm{h}_\text{AF}$. It is noteworthy that Eq.~\eqref{Drude_no_external_field} does not depend on the antiferromagnetic molecular field and therefore it is useful to evaluate the \textit{sublattice-dependent} ASOC. Thus, the NLC provides a way to experimentally deduce the sublattice-dependent ASOC, although it was called "hidden spin polarization"~\cite{Zhang2014HiddenSpin,Gotlieb2018HiddenSpinInCuprate} because it is hard to be measured. Equivalence of $\sigma^{z;xy}_\text{D} =\sigma^{x;yz}_\text{D} =\sigma^{y;zx}_\text{D}$ holds independent of parameters and it can be tested by experiments. Numerical calculations of Eq.~\eqref{drude} are consistent with the above-mentioned symmetry argument and analytic formula as shown in Appendix~\ref{App_Sec_NLC_no_magnetic_field}. A typical value of the nonlinear Hall response is obtained as $\sigma^{z;xy}_\text{D}/[(\sigma^{xx})^2 \sigma^{zz}] \sim \mr{10^{-17}}{[A^{-2} \cdot V \cdot m^3]}$ and it is much larger than the experimental value of bilayer WTe$_2$, $\sigma^{y;xx}/(\sigma^{xx})^3 \sim \mr{10^{-19}}{[A^{-2} \cdot V \cdot m^3]}$~\cite{Ma2019BCD_experiment_WTe2}. Because the Drude term is more divergent with respect to $\tau$ than the BCD term, we may see a giant nonlinear Hall response in the \PT{}-symmetric antiferromagnet.

The NLC is a useful quantity not only to evaluate the sublattice-dependent ASOC but also to detect domain states in antiferromagnetic metals~\cite{Watanabe2018grouptheoretical}. Indeed, the sign of the NLC depends on the antiferromagnetic domain and hence it may promote developments in the antiferromagnetic spintronics~\cite{Jungwirth2016,Manchon2019spin-orbit-torque_review}. In fact, the read-out of antiferromagnetic domains has been successfully demonstrated by making use of the NLC~\cite{Godinho2018AFM_reading}. For \bma{} and related materials listed in Ref.~\cite{Watanabe2018grouptheoretical}, the nonlinear Hall effect can be used to identify antiferromagnetic domain states. So far we considered intrinsic contributions. We have shown that the extrinsic contributions from impurity scattering are suppressed due to the  preserved \PT{}-symmetry, and therefore, they are not relevant to the above discussions.

\subsection{Nematicity-assisted dichroism}\label{Sec_nematicity_assisted}

In the absence of the external field, \bma{}-type magnetic materials do not show the longitudinal NLC along the high symmetry axes, namely, $\sigma^{\mu;\mu\mu}=0$. Below, we show that the longitudinal NLC can be induced by magnetic fields. Since the BCD term contributes to only the transverse response, we have only to consider the Drude term. Generally speaking, to obtain a finite longitudinal electronic dichroism, the system is required to possess an anti-symmetric dispersion such as $k_\mu^3$ or higher-order one. According to the group-theoretical classification, the `polarization' in the momentum-space denoted by $k_\mu$ may share the same symmetry as $k_\mu^3$~\cite{hikaruwatanabe2017,Watanabe2018grouptheoretical}. Thus, the momentum-space polarization is a key to realize the longitudinal dichroism.

In \bma{} and related materials, the momentum-space polarization can be induced by the nematicity. We can understand this by the discussion of the magnetopiezoelectric effect~\cite{Varjas2016,hikaruwatanabe2017,Shiomi2019EuMnBi2_MPE,Shiomi2019CaMn2Bi2_MPE,Shiomi2020}. A magnetopiezoelectric effect means that the planer (electronic) nematicity is induced by the out-of-plane electric current. That is written as
		\begin{equation}
		\varepsilon^{xy} = e^{xy;z} J^z,
		\end{equation}
where $\varepsilon^{\mu\nu}$ represents the strain tensor.  It was experimentally discovered in EuMnBi$_2$~\cite{Shiomi2019EuMnBi2_MPE,Shiomi2020} and CaMn$_2$Bi$_2$~\cite{Shiomi2019CaMn2Bi2_MPE} in accordance with theoretical prediction. The response is derived from the anti-symmetrically distorted Fermi surface and hence realizable in the odd-parity magnetic multipole systems. Similar to the conventional piezoelectric effect, we may expect an inverse effect. Given the in-plane nematic order or strain, the system should obtain the momentum-space polarization $P^{\,k_z}$ whose symmetry is the same as the electric current $J^z$,
		\begin{equation}
		P^{\,k_z} = \tilde{e}^{z;xy} \varepsilon^{xy}.
		\end{equation}
 Accordingly, the longitudinal dichroism $\sigma^{z;zz}$ is allowed. Thus, nematicity-assisted dichroism which is unique to the odd-parity magnetic multipole systems is implied. 

The nematicity can be induced by the magnetic field through the spin-orbit coupling. In the model for \bma{} the sublattice-dependent ASOC plays an essential role. By $\bm{h} \ne 0$, the energy spectrum of the lower bands $E_{\bm{k}}^-$ in Eq.~\eqref{energyspectrum_no_magnetic_field} is modified as
		\begin{equation}
		E_{\bm{k}}^- = \epsilon (\bm{k}) - \sqrt{V_{\rm AB}(\bm{k})^2  + \bm{g}(\bm{k})^2+ \bm{h}^2 \pm 2 |\lambda|  }, \label{energyspectrum_with_magnetic_field}
		\end{equation}
where $\lambda^2 = V_{\rm AB}(\bm{k})^2\,\bm{h}^2+ \left[ \bm{g} (\bm{k})\cdot \bm{h}\right]^2 $. The magnetic field not only lifts the Kramers degeneracy but also causes the nematicity through the coupling $\left[ \bm{g}_0 (\bm{k}) \cdot \bm{h}\right]^2$ in $\lambda$, although linear terms in $\bm{h}$ are canceled out between sublattices in sharp contrast to acentric systems studied before~\cite{Ideue2017}. For \bma{} with Dresselhaus-type staggered ASOC~\cite{Manchon2019spin-orbit-torque_review}, the nematicity denoted by $\varepsilon^{xy}$ is maximally induced by the magnetic field $\bm{h}$ parallel to $[110]$ or $[1\bar{1}0]$.

\begin{figure}[htbp]
		\centering 
		\includegraphics[width=75mm,clip]{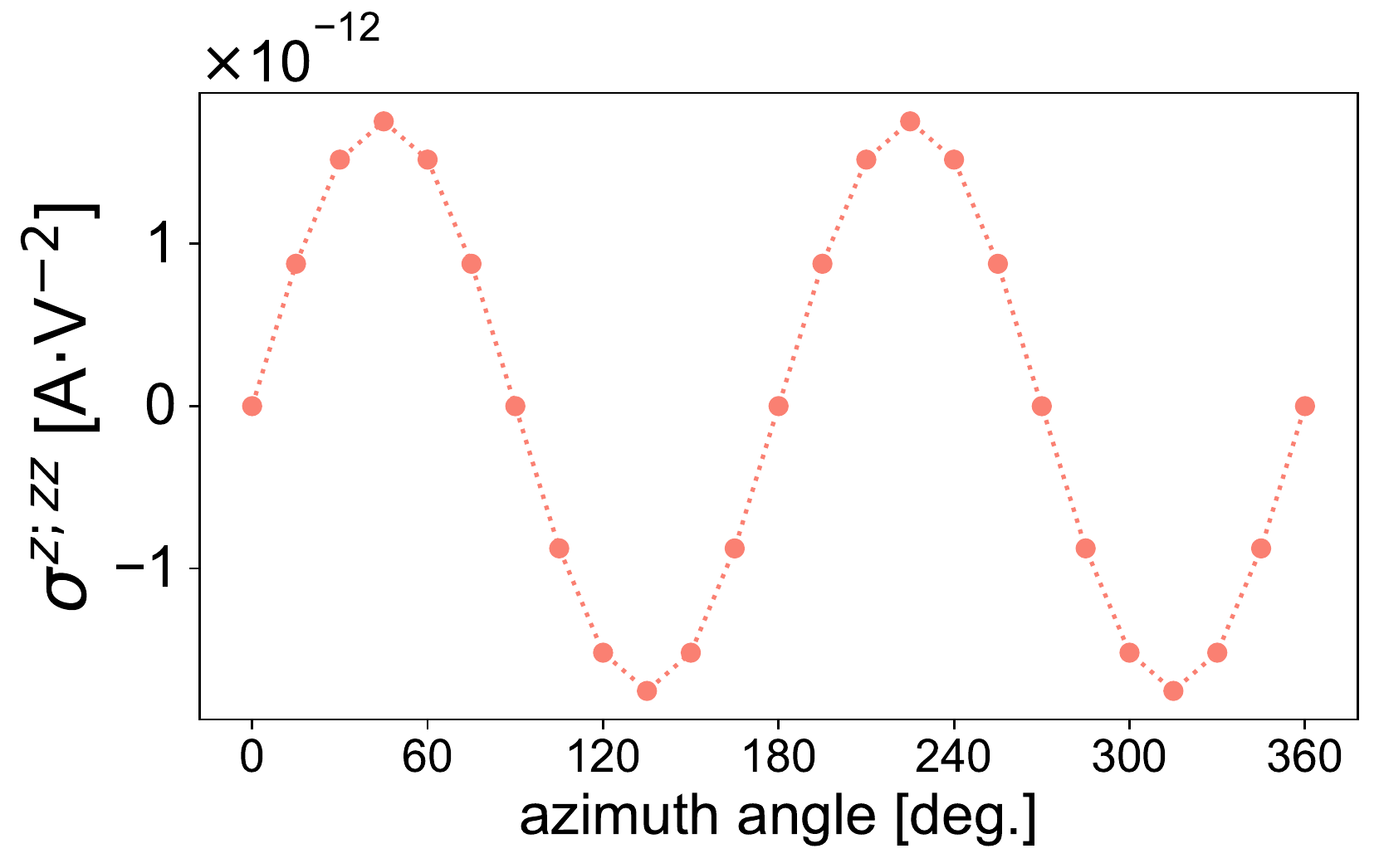}
		\caption{Drude term of a longitudinal NLC $\sigma^{z;zz}_\text{D}$ as a function of the azimuthal angle of external magnetic fields $\bm{h}=h(\cos \phi, \sin \phi, 0)$. Strength of the magnetic field $h=0.01$, temperature $T=0.01$, chemical potential $\mu = -0.5$, relaxation time $\gamma^{-1}=10^3$, and Brillouin zone mesh $N = 135^3$ are adopted. The other parameters and adopted energy scale are described in Appendix~\ref{App_Sec_calc_NLC_Mn_magnet}.}
		\label{Fig_drude_nematicity_assisted_azimuth_dependence} 
		\end{figure}

We expect nematicity-assisted dichroism in \bma{} under the magnetic field $\bm{h} \parallel [110]$ from the above discussions. In numerically calculated NLC $\sigma_\text{D}^{z;zz}$ with rotating the magnetic field in the azimuthal plane, the dichroism with {\it two-fold field-angle dependence} is clearly seen (Fig.~\ref{Fig_drude_nematicity_assisted_azimuth_dependence}). In this case the magnetic field is a \textit{bipolar field} rather than a vector field, in sharp contrast to the magnetic Drude term for which the observed field-angle dependence is one-fold~\cite{Rikken2001magnetochiral_anisotropy,Rikken2005magnetoelectric_anisotropy,Ideue2017}. Although the field-induced NLC is tiny as evaluated in Appendix~\ref{App_Sec_nematic_assisted_NLC}, it was actually detected in a recent experiment for \bma{}~\cite{KimataPrivate}.

\subsection{Magnetic ASOC and Berry curvature dipole}\label{Sec_magneticASOC_BCD}

Now we consider the counterpart of the magnetic Drude term~\cite{Ideue2017}, that is the magnetic BCD term. The \PT{}-symmetry ensures Kramers doublet at each momentum $\bm{k}$, and Berry curvature is completely canceled in the odd-parity magnetic multipole systems. The doublet, however, should be split when the \PT{}-symmetry is broken by the external magnetic field. Let us consider \bma{}-type magnet under the magnetic field $\bm{h}=h_z \hat{z}$ for an example. Then, while the total Berry curvature $\int d\bm{k} \,\Omega^z$ is trivially induced, the BCD also emerges. Using the allowed symmetry operations, the induced BCD is identified as
		\begin{equation}
		\mathcal{D}^{\,xy} = \mathcal{D}^{\,yx}. \label{induced_BCD_in_BMA_with_zField}
		\end{equation}

\begin{figure}[htbp]
		\centering 
		\includegraphics[width=75mm,clip]{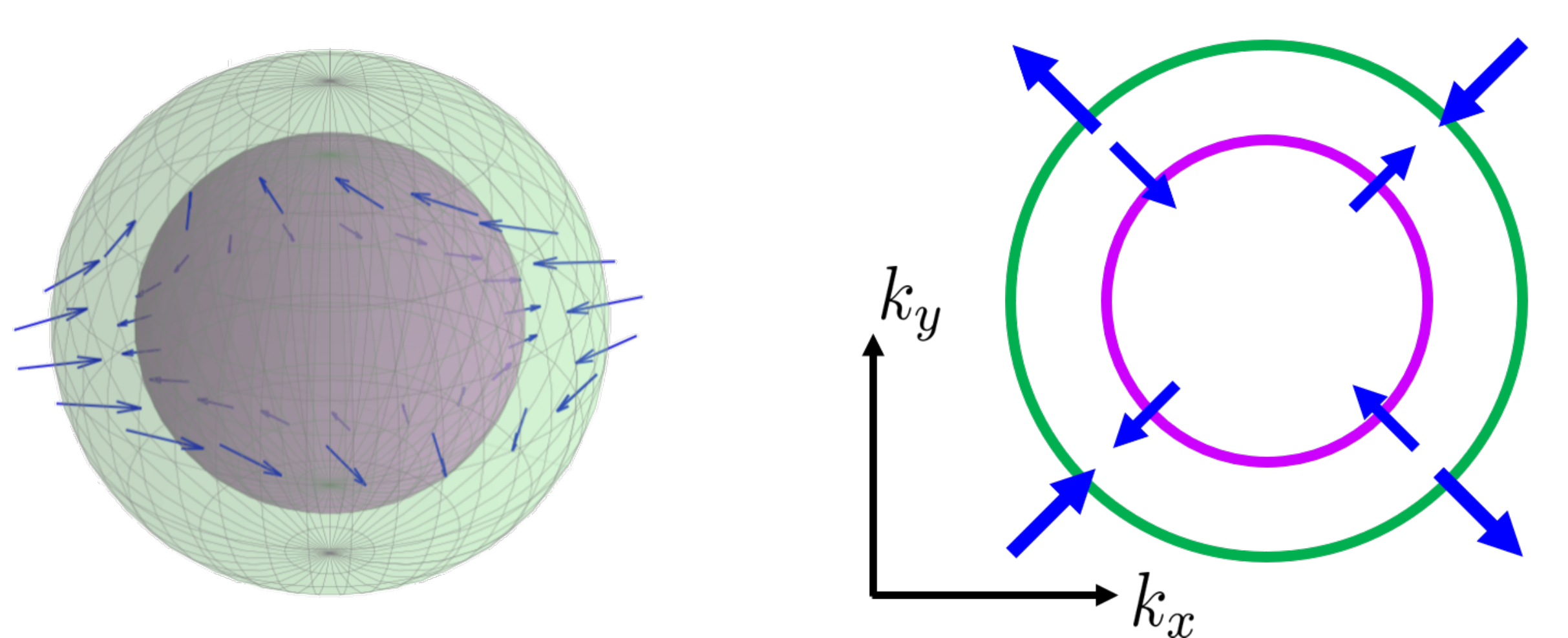}
		\caption{Mechanism of the magnetic ASOC and field-induced BCD. The blue-colored arrows denote the spin-polarization or Berry curvature at each $\bm{k}$. (Left panel) A magnetic field along the $z$-axis splits the Fermi surface depending on the antiferromagnetic molecular field $\bm{h}_\text{AF}$. (Right panel) The split Fermi surface is viewed in the $xy$-plane which indicates the Dresselhaus-type ASOC and BCD.}
		\label{Fig_magnetic_asoc} 
		\end{figure}

Because the BCD has the same symmetry as the ASOC~\cite{Manchon2019spin-orbit-torque_review}, emergence of one indicates the presence of the other.
Therefore, the field-induced BCD is understood by discussing magnetically-induced ASOC in the following way. Although the sublattice-dependent ASOC is compensated with $\bm{h}=0$, combination of the staggered exchange spitting $\bm{h}_\text{AF}\cdot \bm{\sigma}~\tau_z$ and uniform Zeeman field $\bm{h} \cdot \bm{\sigma}~\tau_0$ leads to imbalance between the sublattices without Brillouin zone folding (Fig.~\ref{Fig_magnetic_asoc}). One of sublattices obtains an increased carrier density, and consequently the sublattice-dependent ASOC is not compensated. The emergent ASOC has distinct properties compared to the conventional crystal ASOC since the former originates solely from the magnetic effects. We therefore name this field-induced ASOC `magnetic ASOC'. Interestingly, the magnetic ASOC is tunable by external magnetic fields. Thus, the concept of magnetic ASOC may be useful to design spin-momentum locking in more controllable way than the crystal ASOC which is determined by the crystal structure~\cite{magASOC}. In the model for \bma{} the magnetic ASOC and BCD with the same symmetry as Eq.~\eqref{induced_BCD_in_BMA_with_zField} are actually obtained. 

\begin{figure}[htbp]
		\centering 
		\includegraphics[width=75mm,clip]{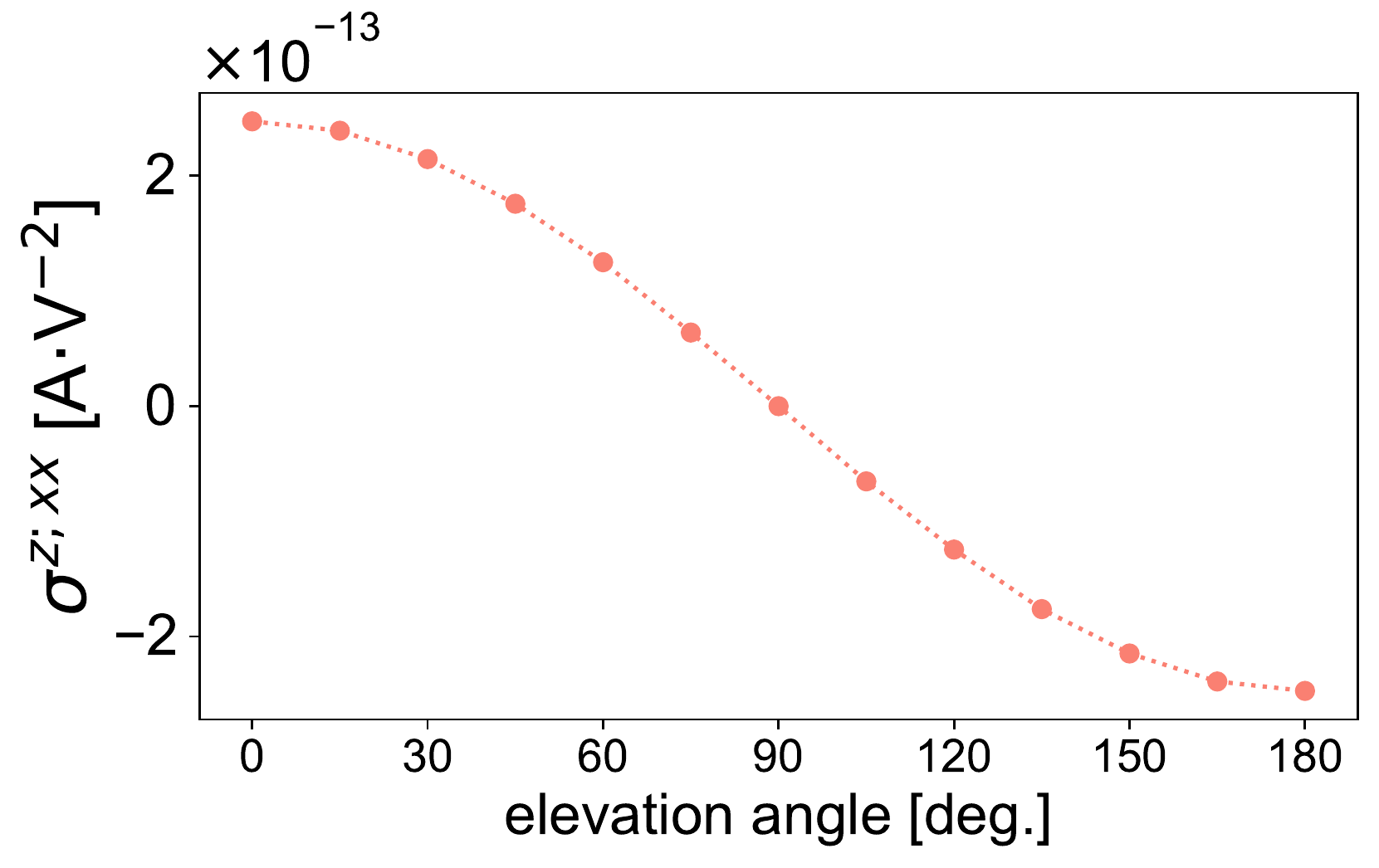}
		\caption{BCD term of a nonlinear Hall conductivity $\sigma^{z;xx}_\text{BCD}$ as a function of the elevation angle of external magnetic fields $\bm{h}=h(\sin \theta, 0, \cos \theta)$. Parameters and unit are the same as Fig.~\ref{Fig_drude_nematicity_assisted_azimuth_dependence}.}
		\label{Fig_BCD_with_z_field_elevation_dependence} 
		\end{figure}

The field-induced BCD allows nonlinear Hall conductivity in accordance with Eq.~\eqref{BCD_term}, which satisfies the relation 
		\begin{equation}
		\sigma_\text{BCD}^{z;xx} = -\sigma_\text{BCD}^{z;yy}= -2\sigma_\text{BCD}^{x;xz}= 2\sigma_\text{BCD}^{y;yz}.
		\label{nonlinear_Hall_by_BCD}
		\end{equation}
For example, we show the numerical result for $\sigma_\mathrm{BCD}^{z;xx}$ in Fig.~\ref{Fig_BCD_with_z_field_elevation_dependence}, which reveals the dependence on the elevation angle of $\bm{h}$. The induced BCD is inverted when the external field is flipped. Therefore, the field-angle dependence is one-fold in contrast to the nematicity-assisted dichroism (Fig.~\ref{Fig_drude_nematicity_assisted_azimuth_dependence}).

Finally, we comment on a linear Hall response. Because the systems under the external magnetic field possess neither the \T{}- nor \PT{}-symmetry, a linear Hall response is also allowed. This is in contrast to the previously studied acentric systems~\cite{Moore2010,Sodemann2015,Xu2018BCD_switchable,Ma2019BCD_experiment_WTe2} where the linear Hall response is forbidden because of the \T{}-symmetry. However, the nonlinear Hall response can be distinguished from the linear one by symmetry. For example, the NLC, $\sigma_\text{BCD}^{z;xx}$ and $\sigma_\text{BCD}^{z;yy}$, in Eq.~\eqref{nonlinear_Hall_by_BCD} represents the Hall response for which the linear response is forbidden.

\section{Conclusion and Discussions}

This work presents symmetry classification of the second-order NLC, and explores the NLC of odd-parity magnetic multipole systems. The Drude term gives rise to a giant nonlinear Hall conductivity at zero magnetic field, and provides an experimental tool for a probe of the sublattice-dependent ASOC. Thus, the hidden spin polarization in centrosymmetric crystals can be clarified. It also enables us to elucidate domain states in antiferromagnetic metals, and hence the NLC will be useful in the field of antiferromagnetic spintronics. Interestingly, the NLC induced by magnetic fields is significantly different from those studied in previous works. 
We clarified the nematicity-assisted dichroism and the BCD-induced NLC due to the magnetic ASOC. 

In accordance with our theoretical result, a recent experimental study actually detected nematicity-assisted electric dichroism under the magnetic field~\cite{KimataPrivate}. 
We believe that further studies of the nonlinear response in parity-violated magnetic systems will be motivated by our work.

\textbf{Acknowledgments}---

The authors are grateful to  A.~Shitade, A.~Daido,~Y. Michishita, M.~Kimata, and R.~Toshio for valuable comments and discussions. Especially, the authors thank M.~Kimata for providing experimental data and motivating this work. This work is supported by a Grant-in-Aid for Scientific Research on Innovative Areas ``J-Physics'' (Grant No.~JP15H05884) and ``Topological Materials Science'' (Grant No.~JP16H00991,~No,~JP18H04225) from the Japan Society for the Promotion of Science (JSPS), and by JSPS KAKENHI (Grant No.~JP15K05164, No.~JP15H05745, and No.~JP18H01178). H.W. is a JSPS research fellow and supported by JSPS KAKENHI (Grant No.~18J23115).

\appendix

\section{Derivation of nonlinear conductivity} \label{App_Sec_derivation_NLC}

We reproduce the expression for second-order electric conductivity~\cite{Sipe1993,Aversa1995,Sipe2000secondorder}. We take the unit $\hbar =1$ below. In general, we have several choices of gauging to introduce an electric field. In an spatially uniform electric field, the Hamiltonian is modified by introducing the vector potential $\bm{A} (t)$ into the canonical momentum $\bm{p}$ as
		\begin{equation}
 		\bm{p} \rightarrow \bm{p} -q \bm{A}(t),\label{velocity_gauge_minimal_coupling}
 		\end{equation}
where $q$ is the charge of carriers and the electric field is obtained as $\bm{E} (t) = - \partial_t \bm{A} (t)$. This choice is called the velocity gauge. On the other hand, the electric field can be taken into account in the Hamiltonian by including 
		\begin{equation}
 		H_\text{E} =  - q \bm{r}\cdot \bm{E} (t), \label{dipole_Hamiltonian}
 		\end{equation} 	
where $\bm{r}$ is the position operator. $H_\text{E}$ is called the dipole Hamiltonian and the gauge choice is called the length gauge. These two choices should give equivalent results because of the gauge invariance~\cite{Ventura2017,Passos2018}. 

The dipole Hamiltonian breaks the translational symmetry in crystals. Therefore, it seems that the Bloch states $\ket{\psi_{\bm{k}a}} = \exp{(i\bm{k}\cdot \hat{\bm{r}})} \ket{u_a (\bm{k})}$ labeled by the crystal momentum $\bm{k}$ and the band index $a$ are not good basis for the total Hamiltonian. However, making use of the Blount's prescription~\cite{Blount1962}, the position operator in the Bloch representation $\bm{r}_{\bm{k}}$  is given by
		\begin{align}
		 \bra{\psi_{\bm{k} a} } \bm{r} \ket{\psi_{\bm{k}' b}} 
			 &=  \delta \left( \bm{k} -\bm{k}' \right) \left[  i \partial_{\bm{k}} \delta_{ab} + \bm{\xi}_{ab}  \right],\\
			 &=  \delta \left( \bm{k} -\bm{k}' \right) \left[ \bm{r}_{\bm{k}} \right]_{ab}.  \label{position_operator}
		\end{align}
We define the Berry connection $\bm{\xi}_{ab} = i \Braket{u_a (\bm{k}) | \partial_{\bm{k}}u_b (\bm{k})}$. The length gauge is adopted in the following calculations.	

\subsection{Density matrix formalism}

Following the Sipe's seminal work~\cite{Sipe1993,Aversa1995,Sipe2000secondorder} and subsequent theoretical studies~\cite{Ventura2017,MatsyshynSodemann2019}, we derive the nonlinear conductivity (NLC) based on the density matrix formalism. Time evolution of the density matrix operator $P(t) = e^{- H(t)/(k_\text{B}T) } /\trace{e^{- H(t)/(k_\text{B}T) }} $ is described by the von-Neumann equation,
		\begin{equation}
		i\partial_t P(t) = [ H(t), P(t) ].\label{von_Neumann_eq}
		\end{equation}
For convenience in the perturbative calculations, we introduce the reduced density matrix,
		\begin{equation}
		\rho_{\bm{k},a b} (t) =  \mathrm{Tr}[ c^\dagger_{\bm{k}b} c_{\bm{k}a} P (t)],
		\end{equation}
where $c_{\bm{k}a}$ is the annihilation operator of the Bloch state $\ket{\psi_{\bm{k}a}}$. In the following, momentum dependence of the reduced density matrix $\rho_{\bm{k}, ab}(t)$ is implicit unless otherwise mentioned. The Hamiltonian consists of the non-perturbative part $H_0$ and the dipole Hamiltonian $H_\text{E} (t)$ in the Schr\"{o}dinger picture. The Bloch state satisfies the equation $H_0 (\bm{k}) \ket{u_{a} (\bm{k})} = \epsilon_{\bm{k}a} \ket{u_{a} (\bm{k})}$ in which $H_0 (\bm{k})$ is the Bloch representation of $H_0$. Thus, Eq.~\eqref{von_Neumann_eq} is recast as
		\begin{equation}
		i \partial_t \rho_{ab} (t) -\left( \epsilon_{\bm{k}a} - \epsilon_{\bm{k}b} \right) \rho_{ab}(t) = -q E_\mu [ r^\mu_{\bm{k}} , \rho  \left( t \right) ]_{ab}.
		\end{equation}
Using the Fourier transformation defined as
		\begin{equation}
		\rho_{ab}\left( t \right) = \int \frac{d\omega}{2\pi} e^{-i\omega t} \rho_{ab} \left( \omega \right),\label{density_matrix_time_fourier_convention}
 		\end{equation}
we obtain 
		\begin{equation}
		\left( \omega -\epsilon_{ab }  \right) \rho_{ab} \left( \omega \right) = -q  \int  \frac{d\Omega}{2\pi} E^\mu \left( \Omega \right)  [ r^\mu_{\bm{k}} , \rho \left( \omega -\Omega \right) ]_{ab}, 
		\end{equation}
where $\epsilon_{ab} = \epsilon_{\bm{k}a} - \epsilon_{\bm{k}b}$. We expand the reduced density matrix $\rho= \sum_n \rho^{(n)}$ by powers of the electric field $\rho^{(n)} = O(|\bm{E}|^n)$, and obtain the recursion formula for the density matrix,
		\begin{equation}
		\left( \omega -\epsilon_{ab }  \right) \rho^{(n+1)}_{ab} \left( \omega \right) = -q  \int  \frac{d\Omega}{2\pi} E^\mu \left( \Omega \right)  [ r^\mu_{\bm{k}}, \rho^{(n)} \left( \omega -\Omega \right) ]_{ab}.
		\end{equation}
In particular, the zero-th order term is obtained as
		\begin{equation}
		\rho^{(0)}_{ab} (\omega)= 2\pi \delta (\omega) f (\epsilon_{\bm{k}a}) \delta_{ab},
		\end{equation}
where $f (\epsilon) = \left[ 1+ \exp{((\epsilon-\mu)/(k_\text{B}T))}\right]^{-1}$ is the Fermi distribution function. The expression is simplified as 
		\begin{equation}
		\rho^{(n+1)}_{ab} \left( \omega \right) = -q \int \frac{d\Omega}{2\pi} d_{ab }^{\,\omega}  E^\mu \left( \Omega \right) [ r^\mu_{\bm{k}} , \rho^{(n)} \left( \omega -\Omega \right) ]_{ab},\label{sequencial_equation_density_matrix}
		\end{equation}
by using the matrix $\hat{d}^{\,\omega}$~\cite{Ventura2017,Passos2018} defined as
		\begin{equation}
		d_{ab}^{\,\omega}  = \frac{1}{\omega - \epsilon_{ab}}.
		\end{equation}
The above derivations are natural extension of the linear response theory~\cite{Kubo1957}.
The position operator in the Bloch representation $r^\mu_{\bm{k}}$ is divided into the diagonal and off-diagonal parts, that is, the first and second terms of Eq.~\eqref{position_operator}. Denoting these two components as $\bm{r}_i$ and $\bm{r}_e$, respectively~\cite{Aversa1995}, the perturbation due to the electric field is classified into the intraband effect $-q \bm{r}_i \cdot \bm{E}$ and interband effect $-q \bm{r}_e \cdot \bm{E}$. We phenomenologically introduce the scattering rate by replacing the matrix $\hat{d}^{\,\omega}$ with
		\begin{equation}
		d_{ab}^{\,\omega}  \rightarrow \frac{1}{\omega + i\gamma - \epsilon_{ab}},
		\end{equation}
where $\gamma$ denotes the scattering rate which is the inverse of the relaxation time $\tau = \gamma^{-1}$~\cite{Passos2018}. This assumption may be satisfied in the presence of the nonmagnetic impurities.

We sequentially obtain corrections to the reduced density matrix $\rho^{(n)}$ ($n>0$). The second-order correction $\rho^{(2)}$ is explicitly written by
		\begin{equation}
		\rho^{(2)}_{ab} (\omega) = \rho^{(ii)}_{ab} (\omega) + \rho^{(ie)}_{ab} (\omega)+ \rho^{(ei)}_{ab} (\omega)+ \rho^{(ee)}_{ab} (\omega),\label{density_matrix_2nd_contribution}
		\end{equation}
where components in the right hand side are labeled by two kinds of the perturbations denoted by intraband (\textit{i}) and interband (\textit{e}) effects. Each component is obtained as
	\begin{widetext}
		\begin{align}
		&\rho^{(ii)}_{ab} (\omega) 
		= (-iq)^2 \int \frac{d\Omega d\Omega'}{(2\pi)^2}  E^\mu (\Omega) E^\nu (\Omega') d_{ab }^{\,\omega} d_{ab}^{\,\omega-\Omega}  \partial_\mu \partial_\nu f(\epsilon_{\bm{k}a}) \times 2\pi \delta_{ab} \delta (\omega - \Omega -\Omega'), \label{density_matrix_2nd_intra_intra_contribution} \\
		&\rho^{(ie)}_{ab} (\omega) 
		= -iq^2 \int \frac{d\Omega d\Omega'}{(2\pi)^2}  E^\mu (\Omega) E^\nu (\Omega') d_{ab }^{\,\omega}  \left[  \partial_\mu \left(  d_{ab }^{\,\omega-\Omega}  f_{ab } \xi^\nu_{ab }  \right) - i \left( \xi^\mu_{aa} - \xi^\mu_{bb}  \right) d_{ab }^{\,\omega-\Omega}  f_{ab } \xi^\nu_{ab }  \right] \times 2\pi  \delta (\omega - \Omega -\Omega'), \\
		&\rho^{(ei)}_{ab} (\omega) 
		= -iq^2 \int \frac{d\Omega d\Omega'}{(2\pi)^2}  E^\mu (\Omega) E^\nu (\Omega') d_{ab }^{\,\omega}   d_{aa }^{\,\omega-\Omega}  \xi^\mu_{ab} \partial_\nu f_{ab }  \times 2\pi  \delta (\omega - \Omega -\Omega'),\\ 
		&\rho^{(ee)}_{ab} (\omega) 
		= q^2 \sum_{c} \int \frac{d\Omega d\Omega'}{(2\pi)^2} E^\mu (\Omega) E^\nu (\Omega') d_{ab }^{\,\omega}  \left[  d_{cb }^{\,\omega-\Omega}  \xi^\mu_{ac} \xi^\nu_{cb}  f_{ bc} - d_{ac}^{\,\omega-\Omega}   \xi^\mu_{cb} \xi^\nu_{ac} f_{ ca}   \right] \times 2\pi  \delta (\omega - \Omega -\Omega'), 
		\end{align}
	\end{widetext}
where $f_{ab } = f(\epsilon_{\bm{k}a}) - f(\epsilon_{\bm{k}b})$. The summation of the repeated Greek indices such as $\mu = x,y,z$ is implicit and $\partial_\mu = \partial / \partial k^\mu$. For perturbative calculations of nonlinear responses, we should respect the intrinsic permutation symmetry between the applied external fields $E^\mu$ and $E^\nu$~\cite{Parker2019}. We hence symmetrize the indices and frequencies of electric fields. For instance, Eq.~\eqref{density_matrix_2nd_intra_intra_contribution} is modified as
	\begin{widetext}
		\begin{equation}
		\rho^{(ii)}_{ab} (\omega) 
		= \frac{(-iq)^2}{2!} \int \frac{d\Omega d\Omega'}{(2\pi)^2}  E^\mu (\Omega) E^\nu (\Omega') d_{ab }^{\,\omega} d_{ab }^{\omega-\Omega}  \partial_\mu \partial_\nu f(\epsilon_{\bm{k}a}) \times 2\pi \delta_{ab} \delta (\omega - \Omega -\Omega') + \left[ \left( \mu, \Omega \right) \leftrightarrow \left( \nu, \Omega' \right) \right].
		\end{equation}
	\end{widetext}

The expectation value of the current density is given by 
		\begin{equation}
		\bm{J}(t) = \text{Tr}[q \bm{v}^{(\text{E})}  P (t)],
		\label{current_expectation_value}
		\end{equation}
where $\bm{v}^{(\text{E})}$ is the velocity operator in the length gauge. In this way, we should express the velocity operator in a given gauge. Starting from the first quantization in the Heisenberg picture, the velocity operator is given by
		\begin{equation}
 		\left[  v^{(\text{E})} (t)\right]^\mu   = \left[ \partial_t r^{(\text{E})}(t)  \right]^\mu = \frac{1}{i} \left[ r^\mu (t), H (t) \right].
 		\end{equation} 
Since the electric field is introduced by taking the dipole Hamiltonian [Eq.~\eqref{dipole_Hamiltonian}] into account, the velocity operator is not modified in the length gauge and should be identical with the unperturbed velocity operator~\cite{Ventura2017}. The unperturbed velocity operator in the Bloch representation is given by
		\begin{equation}
 		\bm{v}_{ab}  \left[ = \bm{v}^{(\text{E})}_{ab} \right]
		= \nabla_{\bm{k}} \epsilon_a \delta_{ab} + i \epsilon_{ab} \bm{\xi}_{ab}.\label{velocity_operator}
 		\end{equation} 
Note that the velocity operator does not coincide with the unperturbed velocity operator in the velocity gauge [Eq.~\eqref{velocity_gauge_minimal_coupling}] because of the non-commutative property between the position operator and the perturbative part of the Hamiltonian arising from Eq.~\eqref{velocity_gauge_minimal_coupling}~\cite{Passos2018,Parker2019}.

\subsection{Second-order nonlinear conductivity}
Here, we derive the second-order NLC by making use of the results in the previous section. The expectation value of the current density proportional to $|\bm{E}|^2$ is
		\begin{align}
		J^{\mu (2)} (\omega) 
			&= \int \frac{d\bm{k}}{\left( 2\pi \right)^d} \sum_{a,b} q v^\mu_{ab} \rho^{(2)}_{ba} (\omega),\\
			&\equiv  \int \frac{d\omega_1d\omega_2}{(2\pi)^2}  \tilde{\sigma}^{\mu;\nu\lambda} (\omega; \omega_1,\omega_2 ) E^\nu \left( \omega_1  \right)E^\lambda \left( \omega_2  \right).
		\end{align}
Since all the components of the reduced density matrix $\rho^{(2)}$ [Eq.~\eqref{density_matrix_2nd_contribution}] have the coefficient $2\pi \delta (\omega - \omega_1 -\omega_2 )$, we take the convention of the second-order NLC tensor $\sigma^{\mu;\nu \lambda}$ as 
		\begin{equation}
 		\tilde{\sigma}^{\mu;\nu \lambda} \left( \omega; \omega_1,\omega_2 \right)= 2\pi \delta (\omega -\omega_1 -\omega_2) ~\sigma^{\mu;\nu \lambda} \left( \omega; \omega_1,\omega_2 \right).
 		\end{equation} 
Substituting the right hand side of Eq.~\eqref{density_matrix_2nd_contribution} for $\rho_{ab}^{(2)} (\omega)$, we express $\sigma^{\mu;\nu \lambda}$ as
		\begin{align}
		\sigma^{\mu;\nu \lambda} 
				&= \sigma^{\mu;\nu \lambda}_\text{D} + \sigma^{\mu;\nu \lambda}_\text{G} + \sigma^{\mu;\nu \lambda}_\text{e}. \label{app_second_order_nonlinear_conductivity_1}
		\end{align}
		
The first term (Drude term) is derived from the component $\rho^{(ii)}$ in Eq.~\eqref{density_matrix_2nd_contribution} and arises from only the intraband effects. The expression is given by  
		\begin{align}
		&\sigma^{\mu;\nu \lambda}_\text{D} \left( \omega; \omega_1,\omega_2 \right)\notag \\
			&= -\frac{q^3}{2}\int \frac{d\bm{k}}{\left( 2\pi \right)^d} \sum_a v^\mu_{aa} d_{aa}^{\,\omega} d_{aa}^{\,\omega_2}  \partial_\nu \partial_\lambda f(\epsilon_{\bm{k}a}) \notag \\
			&~~~~~+ \left[ \left( \nu, \omega_1 \right) \leftrightarrow \left( \lambda, \omega_2 \right) \right],  \\
			&= -\frac{q^3}{2}\int \frac{d\bm{k}}{\left( 2\pi \right)^d} \sum_a \frac{1}{(\omega +i\gamma)(\omega_2 +i\gamma)} \partial_\mu \partial_\nu \partial_\lambda  \epsilon_{\bm{k}a}   f(\epsilon_{\bm{k}a})\notag \\
			&~~~~~+ \left[ \left( \nu, \omega_1 \right) \leftrightarrow \left( \lambda, \omega_2 \right) \right],\label{2nd_order_conductivity_Drude}
		\end{align}
where we use Eq.~\eqref{velocity_operator}. The Drude term can be captured by the conventional Boltzmann's transport theory in which only the intraband effect is semiclassically treated without geometric effects~\cite{Ideue2017}. 

The second term $\sigma^{\mu;\nu \lambda}_\text{G}$ (Geometric term) is derived from the component $\rho^{(ei)}$ and obtained as
		\begin{align}
		&\sigma^{\mu;\nu \lambda}_\text{G} \left( \omega; \omega_1,\omega_2 \right)\notag \\
			&= \frac{q^3}{2}\int \frac{d\bm{k}}{\left( 2\pi \right)^d} \sum_{a\neq b}  d_{ba }^{\,\omega}   d_{aa}^{\,\omega_2} \epsilon_{ab} \xi^\mu_{ab}  \xi^\nu_{ba } \partial_\lambda f_{ba} \notag \\
			&~~~~~+ \left[ \left( \nu, \omega_1 \right) \leftrightarrow \left( \lambda, \omega_2 \right) \right].\label{2nd_order_conductivity_geometric_term}
		\end{align}
The third term $\sigma^{\mu;\nu \lambda}_\text{e}$ due to $\rho^{(ie)}$ and $\rho^{(ee)}$ is written as
	\begin{widetext}		
		\begin{align}
		\sigma^{\mu;\nu \lambda}_\text{e} \left( \omega; \omega_1,\omega_2 \right)
			&= \frac{q^3}{2}\int \frac{d\bm{k}}{\left( 2\pi \right)^d} \sum_{ab} v^\mu_{ab}  d_{ ba}^{\,\omega} \left[ -i    \partial_\nu \left( d_{ba}^{\,\omega_2}  f_{ba} \xi^\lambda_{ba }  \right)  - \left( \xi^\nu_{bb} - \xi^\nu_{aa} \right) d_{ba}^{\,\omega_2}  f_{ba} \xi^\lambda_{ba }\right]\notag \\
			& + v^\mu_{ab}  d_{ ba}^{\,\omega} \left[  \sum_c \left(    d_{ca }^{\,\omega_2}  \xi^\nu_{bc} \xi^\lambda_{ca}  f_{ac} - d_{bc}^{\,\omega_2}   \xi^\nu_{ca} \xi^\lambda_{bc} f_{cb}  \right) \right] +
			 \left[ \left( \nu, \omega_1 \right) \leftrightarrow \left( \lambda, \omega_2 \right) \right].\label{2nd_order_conductivity_interbands}
		\end{align}
	\end{widetext}

Now we obtained the full expression of the second-order NLC. Although the expression seems complicated, it is simplified by making use of the symmetry. In the next section, we clarify the constraints from the basic \T{}- and \PT{}-symmetries.

\subsection{Symmetry constraints on nonlinear conductivity tensor} \label{Sec_symmetry_analysis_2nd_order_conductivity}

It is well known that the crystal symmetries impose strong constraints on physical quantities such as equilibrium properties and transport coefficients~\cite{cracknell2016magnetism}. Furthermore, by combining with expressions derived from microscopic calculations, the symmetry also simplifies physical quantities expressed in the Bloch representation~\cite{Zelezny2017,hikaruwatanabe2017}. We can hence distinguish which intraband effect or interband effect is relevant in a given response function. 

Let us consider the spin polarization induced by the electric field for example. Under the \T{}-symmetry the response coefficients are determined by intraband contributions, whereas the response arises from interband contributions in \PT{}-symmetric systems~\cite{Watanabe2018grouptheoretical}. These responses are hence distinguished and classified as Edelstein effect and magnetoelectric effect in the \T{}-symmetric and \PT{}-symmetric systems, respectively. In the framework of the multipole-based classification, the parity-violating \T{}-/\PT{}-symmetric systems are called odd-parity electric/magnetic multipole systems~\cite{Watanabe2018grouptheoretical,Hayami2018Classification}. Thus, the representation theory of multipole is also useful to associate response functions with symmetry. In a similar manner, we conduct a symmetry analysis of NLC in Eqs.~\eqref{2nd_order_conductivity_Drude},~\eqref{2nd_order_conductivity_geometric_term}, and~\eqref{2nd_order_conductivity_interbands}.

\subsubsection{Drude term}

First, we discuss the Drude term. Equation~\eqref{2nd_order_conductivity_Drude} shows that the second-order Drude conductivity $\sigma_\text{D}$ is determined by the Fermi surface effect as denoted by $\partial f / \partial \epsilon$ and that this term is finite only when the system has an anti-symmetric component in the energy spectrum $\epsilon_{\bm{k}a}$. According to the representation theory of multipole degrees of freedom in solids~\cite{hikaruwatanabe2017,Watanabe2018grouptheoretical,Hayami2018Classification}, the asymmetric dispersion is a striking property of the \PT{}-symmetric odd-parity magnetic multipole systems~\cite{Yanase2014zigzag,Hayami2014h,Hayami2016f,Sumita2016}. The bases of multipoles in the momentum space for such systems are spin-independent and anti-symmetric as
		\begin{equation}
		k_x,~k_xk_yk_z,
		\end{equation}
and these bases imply the anti-symmetric modulation in the energy spectrum of elementary excitations such as electrons and magnons~\cite{Yanase2014zigzag,Hayami2016f}. Thus, we may see a second-order Drude conductivity in odd-parity magnetic multipole systems.

On the other hand, in the \T{}-symmetric odd-parity electric multipole systems, the momentum-space bases are spin-dependent such as 
		\begin{equation}
		k_x \hat{y} - k_y \hat{x},
		\end{equation}
in which $\hat{x}$ and $\hat{y}$ are \T{}-odd pseudo-vectors representing spin polarization, Berry curvature, and so on. These bases represent spin-momentum locking arising from the parity violation~\cite{Manchon2019spin-orbit-torque_review}. Indeed, odd-parity electric multipole systems include familiar noncentrosymmetric crystals. Meanwhile, the spin-independent and anti-symmetric basis does not exist in \T{}-symmetric systems. In fact, the Kramers doublet $\{ \ket{u_{a} (\bm{k})}, \ket{u_{\bar{a}} (-\bm{k})} \}$ protected by the \T{}-symmetry gives rise to the degeneracy between the $\pm \bm{k}$ points as $\epsilon_{\bm{k}a} = \epsilon_{-\bm{k}\bar{a}}$. Therefore, the energy spectrum is symmetric, and the second-order Drude conductivity vanishes in the odd-parity electric multipole systems. However, external magnetic fields may induce the Drude term~\cite{Ideue2017}. 

Summarizing the above-mentioned symmetry analysis, the second-order Drude conductivity is finite if and only if both of the \Pa{} and \T{}-symmetries are violated, and the allowed components due to asymmetric dispersions are indicated by the momentum-space basis of multipoles shown in the classification results~\cite{hikaruwatanabe2017,Watanabe2018grouptheoretical,Hayami2018Classification}. This symmetry requirement is satisfied in the odd-parity magnetic multipole systems with and without external fields~\cite{Watanabe2018grouptheoretical,Hayami2018Classification} as well as in the noncentrosymmetric systems under the external magnetic field~\cite{Rikken2001magnetochiral_anisotropy,Rikken2005magnetoelectric_anisotropy,Tokura2018nonreciprocal_review}. In those systems, the Drude term may give rise to a sizable NLC in the clean limit. Taking the static limit $(\omega, \omega_1, \omega_2 \rightarrow 0)$ in Eq.~\eqref{2nd_order_conductivity_Drude}, the expression is recast as
		\begin{equation}
		\sigma^{\mu;\nu \lambda}_\text{D} \static{} \frac{q^3}{\gamma^2}\int \frac{d\bm{k}}{\left( 2\pi \right)^d} \sum_a \partial_\mu \partial_\nu \partial_\lambda  \epsilon_{\bm{k}a}   f(\epsilon_{\bm{k}a}), 
		\label{app_Drude_term}
		\end{equation}
whose relaxation time dependence is $O(\tau^2)$. Thus, the Drude term is the dominant contribution to the NLC in clean metals.

\subsubsection{Geometric term}

Next, we consider the geometric contribution $\sigma_\text{G}$. In the case of the \PT{}-symmetric systems, the energy spectrum at each $\bm{k}$ has two-fold degeneracy, that is, Kramers doublet. Explicitly denoting the Kramers doublet of Bloch states, 
		\begin{equation}
		\ket{u_{a} (\bm{k})} = \ket{u_{A,\rho} (\bm{k})}, 
		\end{equation}
we introduce the Pauli matrices $\bm{\rho}$ spanned by the Kramers degrees of freedom. Then, a Bloch state is characterized by the index of energy band $A$ and the Kramers degrees of freedom $\rho = \pm$. The transformation property of Kramers doublet can be taken as~\cite{hikaruwatanabe2017}
		\begin{align}
		\mathcal{PT} \ket{u_{A,\rho} (\bm{k})} 
			&= \sum_{\rho'} \ket{u_{A,\rho'} (\bm{k})} (-i\rho_y)_{\rho'\rho},\\
			&= \ket{u_{A,\bar{\rho}} (\bm{k})} (-i\rho_y)_{\bar{\rho}\rho}, \label{Kramers-PT}
		\end{align}
where $\bar{\rho} = - \rho$. Accordingly, matrix elements of the interband Berry connection are transformed as
		\begin{equation}
		\xi^\mu_{ab} (\bm{k}) = \xi^\mu_{A\rho,B\rho'} (\bm{k})= - \sum_{\tau,\tau'}\xi^\mu_{B\tau,A\tau'} (\bm{k}) (-i\rho_y)^\dagger_{\rho'\tau} (-i\rho_y)_{\tau'\rho}.\label{Berry_connection_PT_symmetry_requirement}
		\end{equation}
Of course, the Kramers doublet denoted by $\ket{u_{A,\rho} (\bm{k})}$ with $\rho =\pm$ have the same energy, $\epsilon_{\bm{k}A\rho} = \epsilon_{\bm{k}A\bar{\rho}}$. Hence, we obtain the following relation for  Eq.~\eqref{2nd_order_conductivity_geometric_term} 
		\begin{align}
		&\sum_{a\neq b}  d_{ba }^{\,\omega}    \epsilon_{ab} \xi^\mu_{ab}  \xi^\nu_{ba } \partial_\lambda f_{ba}\notag \\
		&=  \sum_{a \neq b} \sum_{c,d,e,f} d_{ba}^{\,\omega}   \epsilon_{ba} \xi^\mu_{cd}  \xi^\nu_{ef} \partial_\lambda f_{ab }\notag \\
		&~~\times (-i\rho_y)^\dagger_{bc} (-i\rho_y)_{da} (-i\rho_y)^\dagger_{ae} (-i\rho_y)_{fb},\\
		&=  \sum_{\bar{a} \neq \bar{b}} d_{\bar{b}\bar{a} }^{\,\omega}    \epsilon_{ \bar{a}\bar{b}} \xi^\mu_{\bar{b} \bar{a}}  \xi^\nu_{\bar{a} \bar{b} } \partial_\lambda f_{\bar{b} \bar{a}},\\
		&=  \sum_{a \neq b} d_{ba}^{\,\omega}    \epsilon_{ab} \xi^\mu_{ba}  \xi^\nu_{ab } \partial_\lambda f_{ba},		
		\end{align}
in which we use the abbreviated label $\bar{a} = (A\bar{\rho})$. Because of the \PT{}-symmetry, the product of the Berry connections, $\xi^\mu_{ab}\xi^\nu_{ba }$ and $\xi^\mu_{ba}  \xi^\nu_{ab}$, are related to each other in the symmetric way under $\nu\leftrightarrow \lambda$. By using the obtained relations, Eq.~\eqref{2nd_order_conductivity_geometric_term} is simplified as
		\begin{align}
		&\frac{q^3}{2}\int \frac{d\bm{k}}{\left( 2\pi \right)^d} \sum_{a\neq b}  d_{ba }^{\,\omega}   d_{a a }^{\,\omega_2} \epsilon_{ab} \xi^\mu_{ab}  \xi^\mu_{ba } \partial_\nu f_{ba}\notag \\
			&= \frac{q^3}{2}\int \frac{d\bm{k}}{\left( 2\pi \right)^d} \frac{1}{2(\omega_2+i\gamma)}\notag \\
			&\times \left( \sum_{a\neq b}   d_{ba }^{\,\omega}    \epsilon_{ab} \xi^\mu_{ab}  \xi^\nu_{ba } \partial_\lambda f_{ba}  +\sum_{\bar{a}\neq \bar{b}}  d_{ \bar{b}\bar{a} }^{\,\omega}    \epsilon_{ \bar{a}\bar{b}} \xi^\mu_{\bar{b} \bar{a}}  \xi^\nu_{\bar{a} \bar{b} } \partial_\lambda f_{\bar{b}\bar{a} } \right),\label{App_geometric_PT_contraint_eq}\\
			&=  \frac{q^3}{2}\int \frac{d\bm{k}}{\left( 2\pi \right)^d} \frac{1}{2(\omega_2+i\gamma)} \sum_{a\neq b} \left(  d_{ba }^{\,\omega}  + d_{ab }^{\,\omega}  \right) \epsilon_{ab} \xi^\mu_{ab}  \xi^\nu_{ba } \partial_\lambda f_{ba}.
		\end{align}
Taking the static limit, we have
		\begin{align}
		 \frac{d_{ba }^{\,\omega}  + d_{ab }^{\,\omega}}{2(\omega_2+i\gamma)}  
		 		&= \frac{1}{\omega_2+i\gamma} \frac{\omega+i\gamma}{(\omega +i\gamma)^2 - \epsilon_{ab}^2},\\
		 		&\static{} - \frac{1}{\gamma^2 + \epsilon_{ab}^2}.
		\end{align}
We safely take the clean limit ($\gamma \rightarrow 0$) since the expression converges. Finally, we obtain the geometric contribution $\sigma^{\mu;\nu \lambda}_\text{G}$ in the \PT{}-symmetric systems as 
		\begin{align}
		&\sigma^{\mu;\nu \lambda}_\text{G}\rightarrow \sigma^{\mu;\nu \lambda}_\text{e'} \notag \\ 
		&= \frac{q^3}{2}\int \frac{d\bm{k}}{\left( 2\pi \right)^d} \sum_{a\neq b}  \partial_\lambda f_{ab}\frac{\xi^\mu_{ab}  \xi^\nu_{ba }}{\epsilon_{ab}}   + \left[ \nu \leftrightarrow \lambda \right],\\
		&= \frac{q^3}{2}\int \frac{d\bm{k}}{\left( 2\pi \right)^d} \sum_{a\neq b}  \partial_\lambda f(\epsilon_{\bm{k}a}) \left(  \frac{\xi^\mu_{ab}  \xi^\nu_{ba }}{\epsilon_{ab}} + \frac{\xi^\mu_{ba}  \xi^\nu_{ab }}{\epsilon_{ab}}   \right) + \left[ \nu \leftrightarrow \lambda \right],\label{Geometric_term_PT_symmetry}
		\end{align}
which is an intrinsic contribution in the sense that this term is $O(\tau^0)$ and insensitive to the relaxation time. Therefore, we denote the geometric term in the \PT{}-symmetric systems as $\sigma^{\mu;\nu \lambda}_\text{e'}$. In particular, for the two-band Hamiltonian, the products of Berry connections are rewritten by the quantum metric~\cite{Gao2014,gao2019semiclassical}.

Here we consider the \T{}-symmetric systems, where the \T{}-symmetry ensures a relation similar to Eq.~\eqref{Berry_connection_PT_symmetry_requirement},
		\begin{equation}
		\xi^\mu_{A\rho,B\rho'} (\bm{k})= \sum_{\tau,\tau'}\xi^\mu_{B\tau,A\tau'} (-\bm{k}) (-i\rho_y)^\dagger_{\rho'\tau} (-i\rho_y)_{\tau'\rho}.\label{Berry_connection_T_symmetry_requirement}
		\end{equation}
Following the parallel discussion, we obtain the expression of the geometric term in the static limit as
		\begin{align}
		&\sigma^{\mu;\nu \lambda}_\text{G} \notag \\ 
				& = \frac{q^3}{2}\int \frac{d\bm{k}}{\left( 2\pi \right)^d} \frac{1}{2(\omega_2+i\gamma)} \sum_{a\neq b} \left(  d_{ba }^{\,\omega}  - d_{ab }^{\,\omega}  \right) \epsilon_{ab} \xi^\mu_{ab}  \xi^\nu_{ba } \partial_\lambda f_{ba}  \notag \\
				&~~~~~+ \left[ \nu \leftrightarrow \lambda \right],\label{App_geometric_T_contraint_eq} \\
		&\xrightarrow{\text{sta.}}  \sigma^{\mu;\nu \lambda}_\text{BCD}\notag\\
				&= \frac{q^3}{2i\gamma}\int \frac{d\bm{k}}{\left( 2\pi \right)^d} \sum_{a\neq b}  \partial_\lambda f_{ba} \xi^\mu_{ab}  \xi^\nu_{ba }   + \left[ \nu \leftrightarrow \lambda \right],\\
				&=  \frac{i q^3}{2\gamma}\int \frac{d\bm{k}}{\left( 2\pi \right)^d} \sum_{a\neq b}  \partial_\lambda f(\epsilon_{\bm{k}a}) \left(  \xi^\mu_{ab}  \xi^\nu_{ba }   -\xi^\mu_{ba}  \xi^\nu_{ ab }   \right) + \left[ \nu \leftrightarrow \lambda \right], \\
				&=  \frac{q^3}{2\gamma}\int \frac{d\bm{k}}{\left( 2\pi \right)^d} \sum_{a}  \epsilon_{\mu\nu\kappa} \partial_\lambda f(\epsilon_{\bm{k}a})  \Omega^\kappa_a + \left[ \nu \leftrightarrow \lambda \right],\\
				&=  -\frac{q^3}{2\gamma}\int \frac{d\bm{k}}{\left( 2\pi \right)^d} \sum_{a} \epsilon_{\mu\nu\kappa}   f(\epsilon_{\bm{k}a}) \partial_\lambda \Omega^\kappa_a + \left[ \nu \leftrightarrow \lambda \right],
				\label{Geometric_term_T_symmetry}
		\end{align}
where we define the Berry curvature $\Omega^\mu_{a} = \epsilon_{\mu\nu\lambda} \partial_\nu \xi^\lambda_{aa}$ and make use of Eq.~\eqref{Berry_connection_T_symmetry_requirement} in Eq.~\eqref{App_geometric_T_contraint_eq}. The integral in Eq.~\eqref{Geometric_term_T_symmetry},
		\begin{equation}
		\mathcal{D}^{\, \lambda\kappa } = \int \frac{d\bm{k}}{\left( 2\pi \right)^d} \sum_a  f(\epsilon_{\bm{k}a}) \partial_\lambda \Omega^\kappa_a,
		\end{equation}
is called Berry curvature dipole (BCD)~\cite{Sodemann2015}. Therefore, we call the geometric contribution BCD term and it is written as
		\begin{equation}
		\sigma^{\mu;\nu \lambda}_\text{BCD} = -\frac{q^3}{2\gamma} \epsilon_{\mu\nu\kappa}   \mathcal{D}^{\,\lambda\kappa} + \left[ \nu \leftrightarrow \lambda \right],\label{nonlinear_conductivity_Berry_curvature_dipole_term}
		\end{equation}
which has been captured by a semiclassical theory modified to take into account geometrical effects in solids~\cite{Sodemann2015,deyo2009semiclassical,Moore2010}. The BCD term linearly depends on the relaxation time as $O(\tau^1)$ and it is the leading component of NLC in the clean and \T{}-symmetric systems. An important property of the BCD-induced NLC is that the response is always transverse because of Eddington epsilon $\epsilon_{\mu\nu\kappa}$ in Eq.~\eqref{nonlinear_conductivity_Berry_curvature_dipole_term}~\cite{Moore2010,Sodemann2015}.

As shown in Eqs.~\eqref{App_geometric_PT_contraint_eq} and \eqref{App_geometric_T_contraint_eq}, \T{}/\PT{}-symmetry forbids either of symmetric and anti-symmetric part of the Berry connections given by $\xi^\mu_{ab}\xi^\nu_{ba}$. The above classification of the NLC based on the \T{} and \PT{}-symmetries is therefore complementary. Supposing a system without \T{}- and \PT{}-symmetries, all the terms in the above analysis are present and we can use Eqs.~\eqref{Geometric_term_PT_symmetry},~\eqref{Geometric_term_T_symmetry} without modification.

\subsubsection{Summary}

Similarly, we can identify the relaxation time dependence of $\sigma^{\mu;\nu \lambda}_\text{e}$, the last term of Eq.~\eqref{app_second_order_nonlinear_conductivity_1}. Supposing the clean and static limit, $\sigma^{\mu;\nu \lambda}_\text{e}$ is $O (\tau^{-1})$ and $O (\tau^{0})$ in \T{}-symmetric and \PT{}-symmetric systems, respectively. Here, we rewrite the NLC of Eq.~\eqref{app_second_order_nonlinear_conductivity_1} in accordance with the relaxation time dependence,
		\begin{align}
		\sigma^{\mu;\nu \lambda} 
				&= \sigma^{\mu;\nu \lambda}_\text{D} + \sigma^{\mu;\nu \lambda}_\text{G} + \sigma^{\mu;\nu \lambda}_\text{e},\\ 
				&= \sigma^{\mu;\nu \lambda}_\text{D} + \sigma^{\mu;\nu \lambda}_\text{BCD}+ \sigma^{\mu;\nu \lambda}_\text{e'} + \sigma^{\mu;\nu \lambda}_\text{e},\\
				&= \sigma^{\mu;\nu \lambda}_\text{D} + \sigma^{\mu;\nu \lambda}_\text{BCD}+  \sigma^{\mu;\nu \lambda}_\text{int}.
		\end{align}
$\sigma^{\mu;\nu \lambda}_\text{int} = \sigma^{\mu;\nu \lambda}_\text{e'} + \sigma^{\mu;\nu \lambda}_\text{e}$ yields the intrinsic NLC in the clean limit~\cite{Gao2014}. The decomposition is shown in Eq.~\eqref{nonlinear_conductivity}. The symmetry analysis in this section is summarized in Table~\ref{Table_relaxation_time_dependence_2nd_conductivity}. Substituting the electron's charge $q=-e$ in Eqs.~\eqref{app_Drude_term} and \eqref{Geometric_term_T_symmetry}, we obtain Eqs.~\eqref{drude} and~\eqref{BCD_term}.

\section{Calculations of NLC in Mn-based odd-parity magnetic multipole system}\label{App_Sec_calc_NLC_Mn_magnet}
We show the detail of the calculations of the NLC in the odd-parity magnetic multipole systems. In this section, $q= -e$ is taken. 

\subsection{Model Hamiltonian}\label{App_Sec_model_hamiltonian}

		\begin{figure}[htbp] 
		\centering  
		\begin{tabular}{cc}
			\includegraphics[height=40mm]{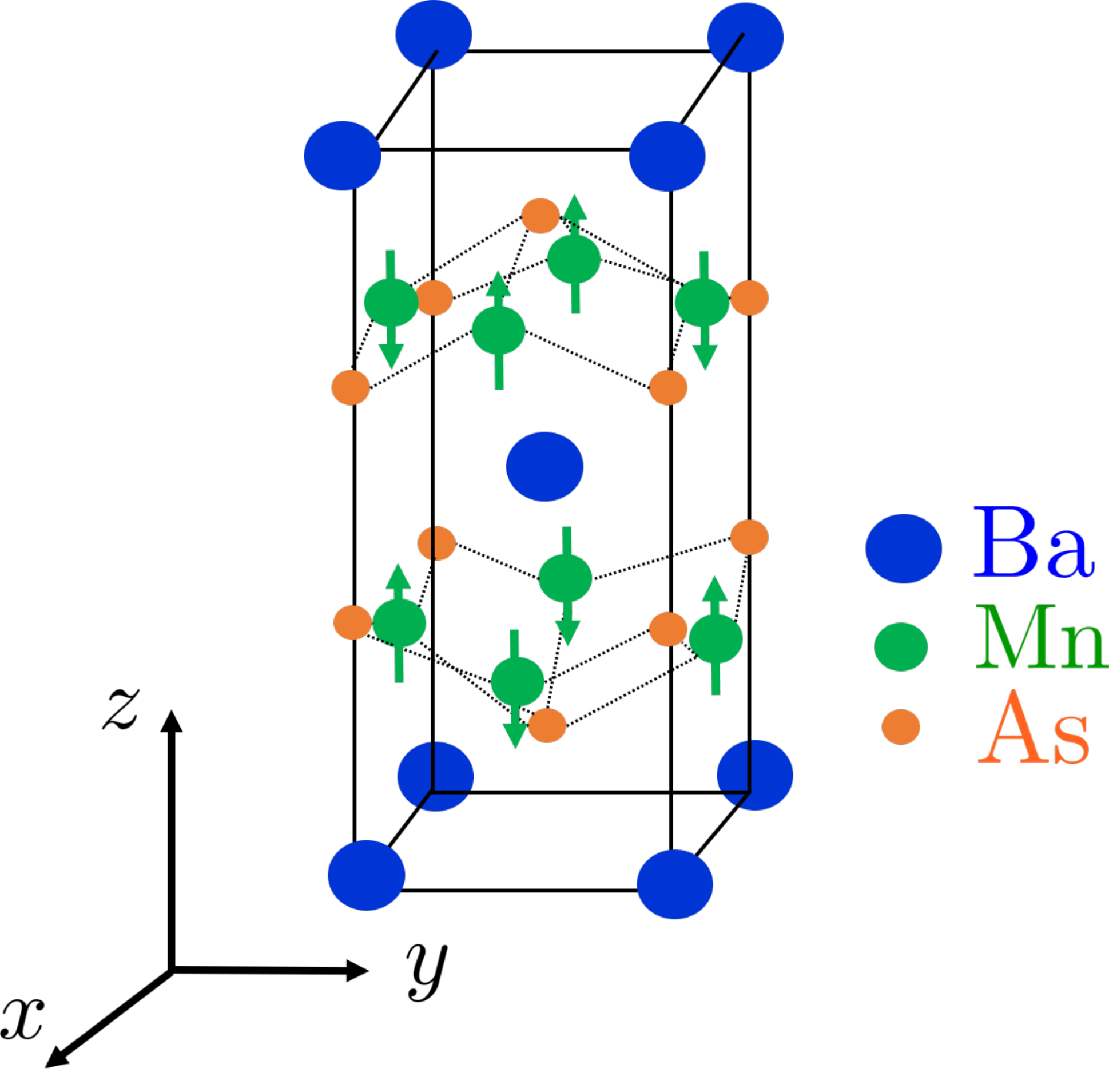} &
			\includegraphics[height=40mm]{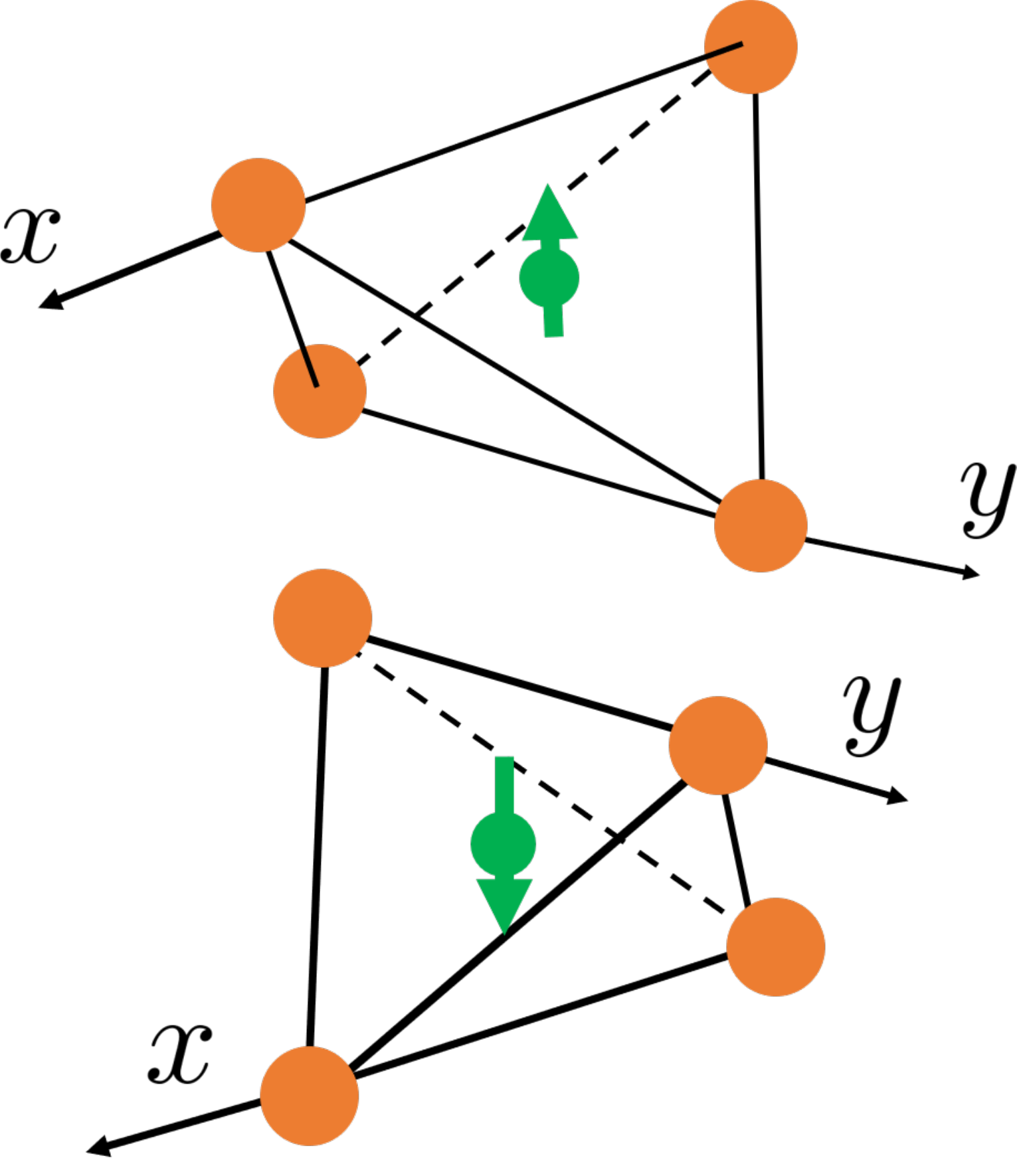} \\
			(a) &
			(b)
			\end{tabular}
		\caption{(a) Crystal and magnetic structures of \bma{}. (b) Two Mn sublattice in \bma{} depicted with surrounding As atoms. Two Mn atoms are not related by the \Pa{}-symmetry in the antiferromagnetic state while they are related in the paramagnetic state.}
		\label{App_Fig_bamnas}  
		\end{figure}

It is known that we have a broad range of candidate materials for the odd-parity magnetic multipole systems, where both of the \Pa{} and \T{}-symmetries are broken while the combined \PT{}-symmetry is preserved~\cite{Spaldin2008a,Watanabe2018grouptheoretical}. In particular, a series of Mn-pnictide compounds are promising candidates and several experimental evidences have been recently reported~\cite{Shiomi2019EuMnBi2_MPE,Shiomi2019CaMn2Bi2_MPE,Shiomi2020}. Thus, we perform microscopic calculations based on the model Hamiltonian for one of the candidate materials, \bma{}~\cite{hikaruwatanabe2017}. The crystal structure of \bma{} is ThCr$_2$Si$_2$-type (space group: $I4/mmm$, No.~139) and Mn atoms are located at the locally-noncentrosymmetric Wyckoff position~\cite{Singh2009BaMn2As2_1,Singh2009BaMn2As2_2}. This system undergoes the G-type antiferromagnetic order and magnetic moments at Mn sites are aligned along the $z$-axis as shown in Fig.~\ref{App_Fig_bamnas}\,(a). Although the magnetic structure is apparently a simple antiferromagnetic, it breaks both of the \Pa{} and \T{}-symmetries instead of the translational symmetry because of the sublattice degree of freedom depicted in Fig.~\ref{App_Fig_bamnas}\,(b). Indeed, it has been shown that the magnetic order is regarded as an odd-parity magnetic multipole order~\cite{hikaruwatanabe2017}. The magnetic structure is denoted by the magnetic point group $I4'/m'm'm$ which has neither \Pa{} nor \T{}-symmetry but respects the \PT{}-symmetry. This compound is semiconducting with narrow energy gap~\cite{Singh2009BaMn2As2_1,Singh2009BaMn2As2_2,Johnston2011}. Doping hole carriers, however, have successfully realized the metallic state of \bma{} without significant modification of antiferromagnetic order and band structure~\cite{Pandey2012BMA_holeDoped,Ramsal2013BMA_MagneticStructure}. Thus, the hole-doped \bma{} is a good example to study the itinerant phenomena such as nonlinear electric transport in odd-parity magnetic multipole ordered systems.

The model Hamiltonian captures the electronic structure observed in the lightly hole-doped \bma{}. The single-orbital model is represented by the Bloch Hamiltonian~\cite{hikaruwatanabe2017},
		\begin{align}
			H(\bm{k}) 
					&=\epsilon(\bm{k}) + \bm{g} (\bm{k}) \cdot \bm{\sigma} \,\tau_z + \bm{h} \cdot \bm{\sigma}   +  V_{\rm AB} (\bm{k}) \tau_x,\label{App_BMA_model_Hamiltonian}\\
					&=\epsilon(\bm{k}) +   \left[\bm{g}_0 (\bm{k}) +  \bm{h}_\text{AF} \right] \cdot \bm{\sigma}\, \tau_z + \bm{h} \cdot \bm{\sigma}   +  V_{\rm AB} (\bm{k}) \tau_x,
		\end{align}
where $\bm{\sigma}$ and $\bm{\tau}$ are Pauli matrices representing the spin and sublattice degrees of freedom, respectively. We introduce the sublattice-dependent anti-symmetric spin-orbit coupling (sASOC), $\bm{g}_0 \cdot \bm{\sigma} \tau_z$, the molecular field of antiferromagnetic order, $\bm{h}_\text{AF} \cdot \bm{\sigma}\, \tau_z$, and the external magnetic field, $\bm{h} \cdot \bm{\sigma}$. In particular, the sASOC characterizes the locally-noncentrosymmetric crystal structure of \bma{} and respects the local symmetry of Mn atoms~\cite{hikaruwatanabe2017}.

The components of the Hamiltonian are given by		
		\begin{align}
		\epsilon (\bm{k})&= -2t_1 \left(	\cos{ k_x  } +\cos{ k_y }	\right)\notag \\
		&~~-8 t_2 \cos{\frac{k_x}{2} } \cos{\frac{k_y}{2} }\cos{\frac{k_z}{2} },\label{intrasub_kinetic}\\
		V_{\rm AB }(\bm{k})&= -4\tilde{t}_1  	\cos{ \frac{k_x}{2}  } \cos{ \frac{k_y}{2} } -2 \tilde{t}_2 \cos{\frac{k_z}{2} },\\
		\bm{g}_0 (\bm{k}) &= \begin{pmatrix}
			\alpha_1  \sin{k_y} + \alpha_2 \cos{\frac{k_x}{2} } \sin{\frac{k_y}{2} }\cos{\frac{k_z}{2} }\\
			\alpha_1  \sin{k_x} + \alpha_2 \sin{\frac{k_x}{2} } \cos{\frac{k_y}{2} }\cos{\frac{k_z}{2} }\\
			\alpha_3 \sin{\frac{k_x}{2} } \sin{\frac{k_y}{2} }\sin{\frac{k_z}{2} }
			\end{pmatrix},\\
		\bm{h}_\text{AF}  &= \left( 0,0,h_\text{AF} \right).
		\end{align}
Hopping parameters $t_i,\tilde{t}_i$ and the sASOC strength $\alpha_i$ are introduced. In accordance with the reported magnetic structure of \bma{}, the  molecular field is taken as $\bm{h}_\text{AF} \parallel z$~\cite{Singh2009BaMn2As2_2,Ramsal2013BMA_MagneticStructure}. The model reproduces the experimentally observed Fermi surface in the lightly hole-doped compounds~\cite{Pandey2012BMA_holeDoped,Zhang2016ARPES}, when the parameters are chosen as
		\begin{equation}
			t_1 = -0.1,~ t_2=-0.05,~\tilde{t}_1=0.05,~\tilde{t}_2=0.01,\label{bamn2as2modelparameter}
		\end{equation}
and
		\begin{equation}
			\alpha_1 = -0.005,~\alpha_2 =0.001,~\alpha_3 = 0.01,~h_\text{AF}=1.\label{bamn2as2modelparameter2}
		\end{equation}
In this work, we adopted these parameters. The remaining parameter $\bm{h}$ is set in the following sections.

\subsection{Nonlinear Hall conductivity at zero magnetic field}\label{App_Sec_NLC_no_magnetic_field}
In this section, we calculate the Drude term of NLC in the absence of the external magnetic field, and hence we take $\bm{h} = \bm{0}$ in Eq.~\eqref{App_BMA_model_Hamiltonian}.

The odd-parity magnetic multipole systems preserve neither the \Pa{}-symmetry nor \T{}-symmetry owing to the parity-violating antiferromagnetic order. The electronic structure, therefore, shows a peculiar property. Diagonalizing the Bloch Hamiltonian of Eq.~\eqref{App_BMA_model_Hamiltonian}, the energy spectrum is obtained as
		\begin{equation}
		E^\pm_{\bm{k}}= \epsilon (\bm{k}) \pm \sqrt{V_{\rm AB}(\bm{k})^2  + \bm{g} (\bm{k})^2 }, \label{App_energyspectrum_under_no_field}
		\end{equation}
where the spectrum has the two-fold degeneracy protected by the \PT{}-symmetry. The coupling between the sASOC and the molecular field written as $\bm{g}_0 (\bm{k}) \cdot \bm{h}_\text{AF}$ gives rise to an anti-symmetric component in the energy dispersion, and thus $E^\pm_{\bm{k}} \ne E^\pm_{-\bm{k}}$.  This anti-symmetric component is reduced to $k_xk_yk_z$ around a time-reversal-invariant momentum, which is consistent with the group-theoretical classification theory~\cite{hikaruwatanabe2017,Watanabe2018grouptheoretical}. Accordingly, the Drude terms of the second-order conductivity, $\sigma_\text{D}^{z;xy}$, $\sigma_\text{D}^{y;xz}$, and $\sigma_\text{D}^{x;yz}$ are allowed in \bma{}, since the Drude term is determined by the anti-symmetric and anharmonic property of the energy spectrum denoted by $\partial_{\mu}\partial_{\nu}\partial_{\lambda} E_{\bm{k}}$ in Eq.~\eqref{2nd_order_conductivity_Drude}. These components are indeed nonlinear Hall conductivity. It is noteworthy that the second-order Drude conductivity is totally-symmetric with respect to the permutation of indices $(\mu,\nu,\lambda)$. Thus, the following relation is satisfied 
		\begin{equation}
		\sigma_\text{D}^{x;yz} = \sigma_\text{D}^{y;xz}=\sigma_\text{D}^{z;xy},
		\end{equation}
in spite of the intrinsic anisotropy of the tetragonal symmetry. We conduct analytical and numerical calculations of the nonlinear Hall conductivity $\sigma_\text{D}^{z;xy}$ below.

To obtain an analytical expression of the second-order Drude conductivity, we assume low carrier density and approximate the energy spectrum up to $O(|\bm{k}|^3)$. The assumption is reasonable for the observed electronic structure in the hole-doped \bma{}~\cite{Pandey2012BMA_holeDoped,Zhang2016ARPES}. The microscopic parameters in Eqs.~\eqref{bamn2as2modelparameter} and~\eqref{bamn2as2modelparameter2} imply
		\begin{equation}
		|\bm{h}_\text{AF}| \gg |t_i|,|\tilde{t}_i|,|\alpha_i|.
		\end{equation}
We hence evaluate the Drude contribution in the static limit as
		\begin{align}
		&\sigma^{z;xy}_\text{D}\notag \\ 
				&\simeq -q^3 \int \frac{d\bm{k}}{\left( 2\pi \right)^3} \sum_{a}  \frac{1}{\gamma^2} \frac{ \left[ \partial_x \partial_y \partial_z \bm{g}_0 (\bm{k}) \right] \cdot \bm{h}_\text{AF}}{|h_\text{AF}|}  f(\epsilon_{\bm{k}a})  \\
				&= -q^3 \int \frac{d\bm{k}}{\left( 2\pi \right)^3} \sum_{a}  \frac{1}{\gamma^2} \frac{\alpha_3 h_\text{AF}}{8\,|h_\text{AF}|}  f(\epsilon_{\bm{k}a}) \\
				&=  -2\times \frac{q^3\alpha_3n }{8\gamma^2}\,\text{sgn}\,(h_\text{AF}),\label{2nd_order_Drude_xyz_low_carrier}
		\end{align}
where $n$ denotes the carrier density. The coefficient $2$ in the last line arises from the two-fold degeneracy due to the \PT{}-symmetry. Substituting the electron's charge $q=-e$, we obtain Eq.~\eqref{Drude_no_external_field}.

\begin{figure}[htbp]
		\centering 
		\includegraphics[width=80mm,clip]{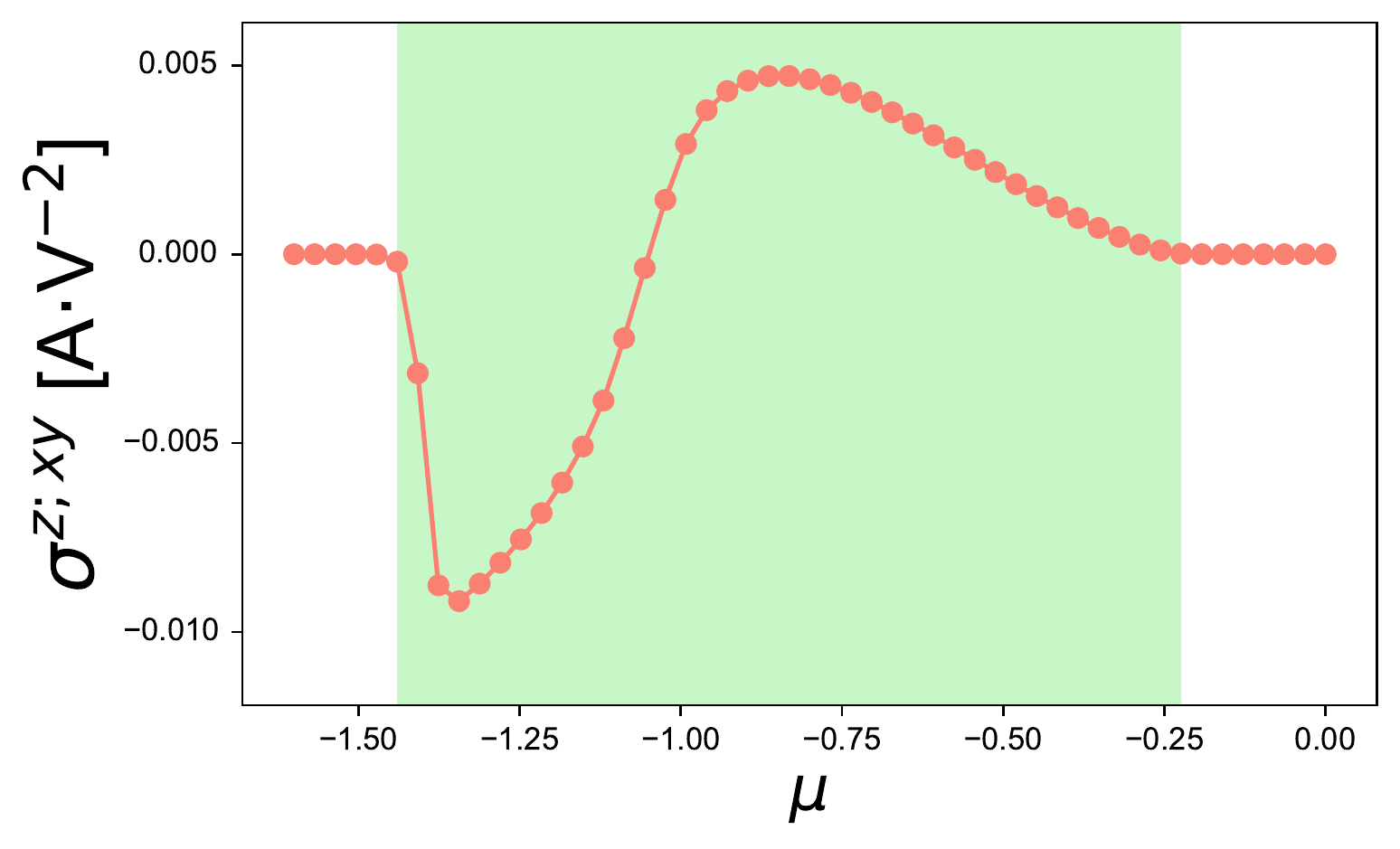}
		\caption{Drude term of nonlinear Hall conductivity $\sigma^{z;xy}_\text{D}$ as a function of the chemical potential $\mu$. Green-colored shaded region indicates the metallic regime where the density-of-states is finite. The parameters are $T=0.01$ and $\gamma^{-1}=1.0\times10^3$. We take $N = 135^3$.}
		\label{Fig_Drude312_no_magnetic_field} 
		\end{figure}

Next, we show the numerical result. Numerical integration for $\bm{k}$ is carried out by adopting $N=L^3$ discretized cells. Taking the chemical potential $\mu$ as a parameter, we plot the Drude term $\sigma^{z;xy}_\text{D}$ in Fig.~\ref{Fig_Drude312_no_magnetic_field}. We confirm that the Drude term is finite only in the metallic state and it is proportional to the square of the relaxation time $\tau^2$. This is the dominant contribution in clean metals because the other allowed contribution, that is, $\sigma_\text{int}$, is $O(\tau^0)$ in the \PT{}-symmetric systems (See Table.~\ref{Table_relaxation_time_dependence_2nd_conductivity}). For a realistic parameter of the hole-doped \bma{}, we should consider a chemical potential near the top of the lower band, that is $\mu \sim -0.5$ in Fig.~\ref{Fig_Drude312_no_magnetic_field}.

Here, we quantitatively estimate the nonlinear Hall conductivity. By taking the energy scale $|t_1| = \mr{1}{eV}$, the Drude term is evaluated to be $\sigma^{z;xy}_\text{D} \sim  10^{-3}$ [A$\cdot$V$^{-2}$]. We also numerically calculate the linear conductivity $\sigma^{\mu\nu}$ with the same parameters (not shown), and the value is estimated to be $\sigma^{xx} \sim 10^{5}$ [A$\cdot$V$^{-1}\cdot$m$^{-1}$] with a lattice constant $a_0 = 10$ [\AA]. Note that the Drude term in the linear conductivity is $O(\tau^1)$ while the second-order Drude conductivity is $O(\tau^2)$. Thus, nonlinear Hall current may not be negligible in spin-orbit coupled clean metals.

Finally, we compare the nonlinear Hall effect calculated for \bma{} with the experimental result of WTe$_2$~\cite{Ma2019BCD_experiment_WTe2}. Recently, the nonlinear Hall effect has been observed in WTe$_2$ at zero magnetic field~\cite{Xu2018BCD_switchable,Ma2019BCD_experiment_WTe2} and many related studies have been reported. For comparison, we consider a similar measurement geometry of the Hall response. In the case of \bma{}, a nonlinear Hall current $J_\text{NL} \parallel $ [001] gives rise to a Hall electric field $E_\text{H}$ in the parallel direction
        \begin{equation}
            E_\text{H} = \frac{J_\text{NL}}{\sigma^{zz}} = \frac{\sigma^{z;xy}_\text{D}}{\sigma^{zz}} E_\text{ext}^2, 
        \end{equation}
where $E_\text{ext}$ is the applied electric field along the [110]-direction. Here $\sigma^{zz}$ represents the out-of-plane linear conductivity, which is numerically estimated to be smaller than $\sigma^{xx}$ by one order of magnitude.  Converting the electric field into the electric current by $E_\text{ext} = J_\text{ext}/\sigma^{xx}$, we obtain the relation between $E_\text{H}$ and $J_\text{ext}$,
        \begin{equation}
            E_\text{H} = \frac{\sigma^{z;xy}_\text{D}}{(\sigma^{xx})^2\sigma^{zz}} J_\text{ext}^2.
        \end{equation}
Assuming the values estimated above, $\sigma^{z;xy}_\text{D} \sim  10^{-3}$ [A$\cdot$V$^{-2}$], $\sigma^{xx} \sim 10^{5}$ [A$\cdot$V$^{-1}\cdot$m$^{-1}$], and $\sigma^{zz} \sim 10^{4}$ [A$\cdot$V$^{-1}\cdot$m$^{-1}$], we obtain
        \begin{equation}
            \frac{\sigma^{z;xy}_\text{D}}{(\sigma^{xx})^2 \sigma^{zz}} \sim \mr{10^{-17}}{[A^{-2} \cdot V \cdot m^3]},
            \label{nonlinear_Hall_effect_BaMn2As2}
        \end{equation}
for a model of \bma{}.

In the case of the bilayer WTe$_2$, the observed voltage drop due to the nonlinear Hall effect is $V_\text{H} \sim \mr{10}{[\mu V]}$ with the applied current $I_\text{ext} \sim \mr{1}{[\mu A]}$~\cite{Ma2019BCD_experiment_WTe2}. Considering the experimental setup in Ref.~\cite{Ma2019BCD_experiment_WTe2} and the thickness of the bilayer WTe$_2$, $d \sim \mr{1}{[nm]}$, the Hall electric field and applied electric current density are evaluated as $E_\text{H} = \mr{10}{[V \cdot m^{-1}]}$ and $J_\text{ext} = \mr{10^{10}}{[A \cdot m^{-2}]}$, respectively. Thus, we estimate 
        \begin{equation}
            \frac{\sigma^{y;xx}}{(\sigma^{xx})^3} \sim \mr{10^{-19}}{[A^{-2} \cdot V \cdot m^3]},
        \label{nonlinear_Hall_effect_WTe2}
        \end{equation}
for the bilayer WTe$_2$. The coordinates of (non)linear conductivity tensors are chosen to be the same as those in  Ref.~\cite{Ma2019BCD_experiment_WTe2}.

Comparing Eq.~\eqref{nonlinear_Hall_effect_BaMn2As2} with Eq.~\eqref{nonlinear_Hall_effect_WTe2}, we expect that the nonlinear Hall response in \bma{} may be much larger than that observed in WTe$_2$. It should be noticed that the nonlinear Hall response of \bma{} is due to the Drude term proportional to the square of the relaxation time $O(\tau^2)$, while that of WTe$_2$ is considered to be determined by the BCD term or extrinsic contributions, which are $O(\tau^{2n+1})$ (See Table~\ref{Table_relaxation_time_dependence_2nd_conductivity} and Ref.~\cite{Du2019}). Hence, cleanness of the sample enhances the nonlinear Hall conductivity in \bma{} much more than it does in WTe$_2$.

\subsection{Nematicity-assisted dichroism}\label{App_Sec_nematic_assisted_NLC}
Next, we investigate the Drude contribution to the NLC induced by the magnetic field. Diagonalizing the effective Hamiltonian in Eq.~\eqref{App_BMA_model_Hamiltonian} with $\bm{h} \ne 0$, we obtain the energy spectrum
		\begin{equation}
		E_{\bm{k}}= \epsilon (\bm{k}) \pm \sqrt{V_{\rm AB}(\bm{k})^2  + \bm{g}(\bm{k})^2+ \bm{h}^2 \pm 2|\lambda|  }, \label{App_energyspectrum_with_magnetic_field}
		\end{equation}
where $\lambda^2 = V_{\rm AB}(\bm{k})^2\,\bm{h}^2+ \left[ \bm{g} (\bm{k})\cdot \bm{h}\right]^2 $. We see that the energy spectrum is modified from Eq.~\eqref{App_energyspectrum_under_no_field}, and the Kramers degeneracy is split due to violation of the \PT{}-symmetry. As we discussed in Sec.~\ref{Sec_nematicity_assisted}, the in-plane magnetic field gives rise to the nematicity in the $xy$-plane. The corresponding term is represented by the coupling between the $g$-vector and the magnetic field given by $\left[ \bm{g}_0 (\bm{k})\cdot \bm{h} \right]^2$. The energy spectrum is therefore symmetric against the transformation $\bm{h} \rightarrow -\bm{h}$.

For comparison, we consider the two-band Hamiltonian for noncentrosymmetric systems
		\begin{equation}
			H(\bm{k}) =
					\tilde{\epsilon}(\bm{k}) + \tilde{\bm{g}} \left( \bm{k} \right) \cdot \bm{\sigma},\label{two_band_Hamiltonian}
		\end{equation}
where the $g$-vector consists of the ASOC and the magnetic field, $\tilde{\bm{g}} (\bm{k})= \tilde{\bm{g}}_0 (\bm{k}) +\bm{h}$. The energy spectrum is obtained as $E =\tilde{\epsilon}(\bm{k}) \pm |\tilde{\bm{g}} (\bm{k})| $. Because of the coupling term $\tilde{\bm{g}}_0 (\bm{k})\cdot \bm{h}$, the dispersion is not invariant when the external field is inverted~\cite{Ideue2017}. Thus, the effect of magnetic field is significantly different between the noncentrosymmetric systems and locally-noncentrosymmetric systems with parity-violating magnetic order.

\begin{figure}[htbp]
			\centering
			\begin{tabular}{c}
			\includegraphics[height=50mm,clip]{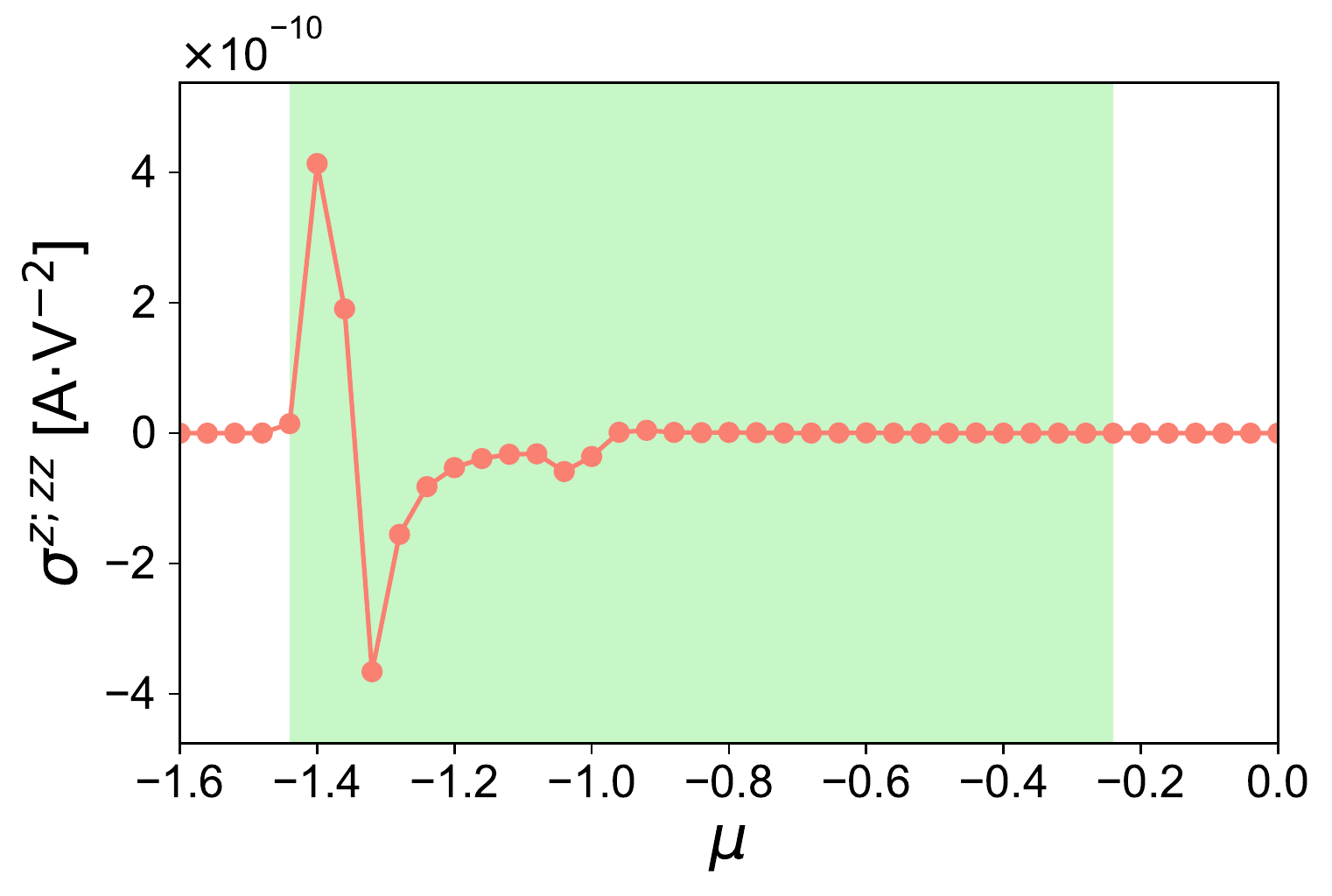} \\
			(a)\\
			\includegraphics[height=50mm,clip]{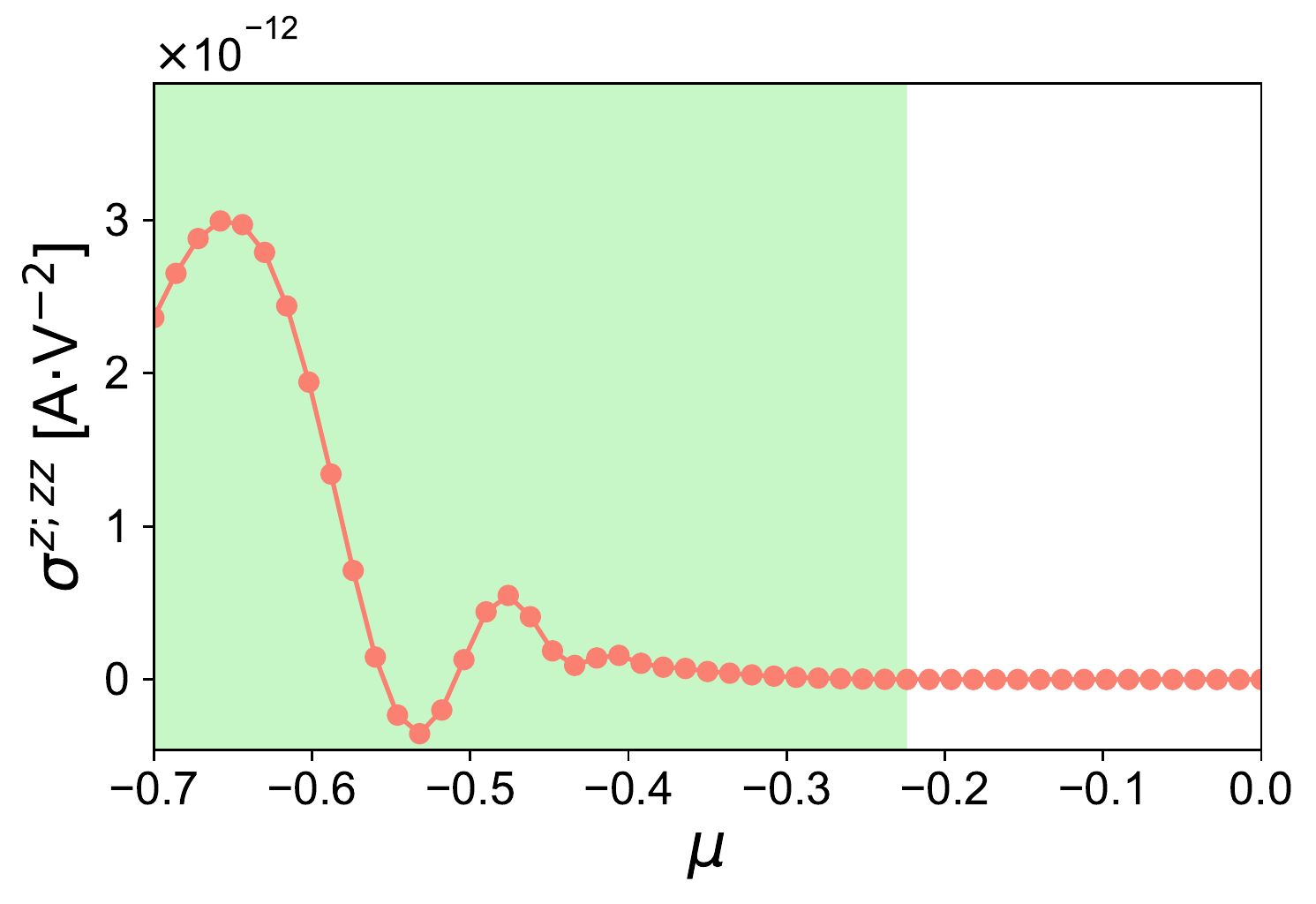} \\
			(b) \\ 
			\end{tabular}
		\caption{(a) Drude term of a longitudinal NLC $\sigma^{z;zz}_\text{D}$ under the external magnetic field as a function of the chemical potential.  (b) Enlarged plot near the top of the band. Green-colored shaded area indicates the metallic regime. The parameters are $T=0.01$, $\gamma^{-1}=1.0\times10^3$, and $h=0.01$. We take $N = 135^3$.}
		\label{Fig_Drude333_with_magnetic_field}
		\end{figure}

As we discussed in Sec.~\ref{Sec_NLC_odd-parity_magnetic_multipole}, a longitudinal NLC $\sigma^{z;zz}_\text{D}$ is induced by the nematicity caused by the in-plane magnetic field. We present a numerical result of $\sigma^{z;zz}_\text{D}$ in Fig.~\ref{Fig_Drude333_with_magnetic_field}. The parameters are the same as Fig.~\ref{Fig_Drude312_no_magnetic_field} except for an additional magnetic field, $\bm{h}\parallel [110]$ with $|\bm{h}|=0.01$. Note that in hole-doped \bma{} a realistic parameter of the chemical potential should be near the top of the lower band ($\mu \sim -0.5$)~\cite{Pandey2012BMA_holeDoped}. In this region, the longitudinal NLC is in the order of $\sigma^{z;zz}_\text{D} \sim 10^{-8}$ which is much smaller than  $\sigma^{z;zz}_\text{D} \sim 10^{-5}$ in the heavily-doped region. This is partly because lightly-doped holes lead to a Fermi pocket near $\bm{k} = \bm{0}$. In the vicinity of $\bm{k} = \bm{0}$ the inter-sublattice hopping term  $V_\text{AB} (\bm{k})$ surpasses the sASOC $\bm{g}_0 (\bm{k})$, and therefore, the effect of the sASOC is not significant~\cite{Maruyama2012}. In fact, when the sign of the hopping integrals $t_i$ in Eq.~\eqref{bamn2as2modelparameter} is inverted, the hole pocket appear around $\bm{k} = (\pi,\pi,\pi)$, where the magnitude of sASOC is comparable to the inter-sublattice hopping energy. Then, the NLC is significantly enhanced. The field-induced dichroism is strongly enhanced in materials having strongly spin-orbit coupled Fermi surfaces.

The magnitude of the nonlinear response is estimated to be $\sigma^{z;zz}_\text{D} \sim  10^{-10}$ [A$\cdot$ V$^{-2}$]. Because a controllable parameter in the experiment is the electric current rather than the electric field, we rewrite the nonlinear response by the electric field generated by the electric current
		\begin{equation}
		 E^z = \rho^{zz} J^z + \rho^{z;zz} (J^z)^2. 
		\end{equation}
The linear term is usually much larger than the second-order term, and hence resistivity tensors are approximated by~\cite{Ideue2017}
		\begin{equation}
		\rho_{zz} = \frac{1}{\sigma^{zz}},~\rho^{z;zz} = \frac{-1}{\sigma^{zz}}\frac{\sigma^{z;zz}}{\left( \sigma^{zz} \right)^2}. 
		\end{equation}
Then, we obtain $\rho^{zz} \sim  \mr{10^{-5}}{[\Omega \cdot m]}$ and $\rho^{z;zz} \sim  -\mr{10^{-25}}{[\Omega \cdot m^3 \cdot A^{-1}]}$ when we assume $\sigma^{zz} \sim 10^{5}$ [A$\cdot$ V$^{-1}\cdot$ m$^{-1}$] and $\sigma^{z;zz}_\text{D} \sim   \mr{10^{-10}}{[A\cdot V^{-2}]}$. Taking $J^z \sim \mr{10^{2}}{[mA\cdot mm^{-2}]}$, the electric field responsible for the linear and nonlinear conductivity are estimated to be $E^{(1)} \sim \mr{10^{0}}{[V\cdot m^{-1}]}$ and $E^{(2)} \sim - \mr{10^{-15}}{[V\cdot m^{-1}]}$, respectively. Although the voltage drop due to the nonlinear conductivity is tiny, it may be detected via the AC measurement~\cite{Ideue2017}. Furthermore, the nonlinear response is enhanced by applying a large electric current in a microscopic sample. A recent experimental study actually detected a longitudinal NLC in \bma{} under the magnetic field~\cite{KimataPrivate}.

\subsection{Nonlinear Hall conductivity due to Berry curvature dipole} \label{App_Sec_symmetry_of_BCD}

\begin{figure*}[htbp]
			\centering
			\begin{tabular}{ccc}
			\includegraphics[height=30mm]{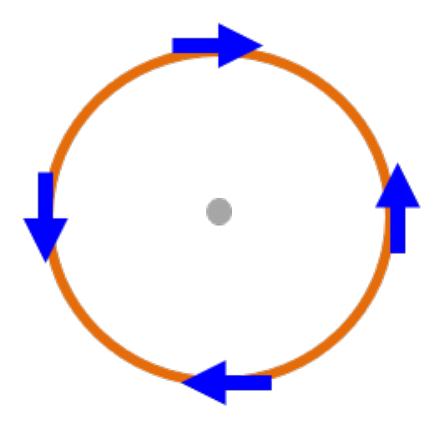} &
			\includegraphics[height=30mm]{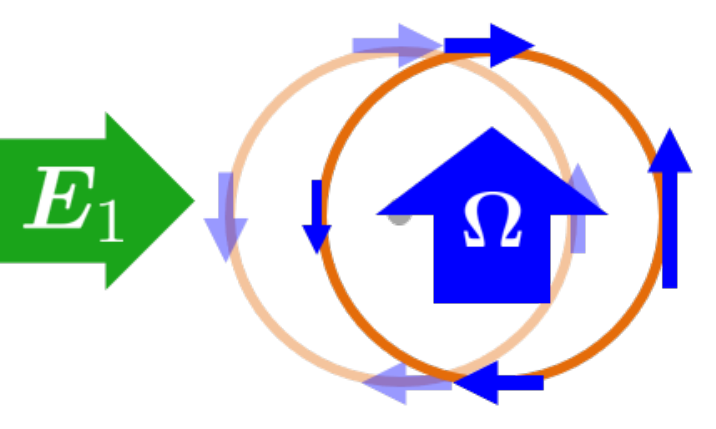}&
			\includegraphics[height=30mm]{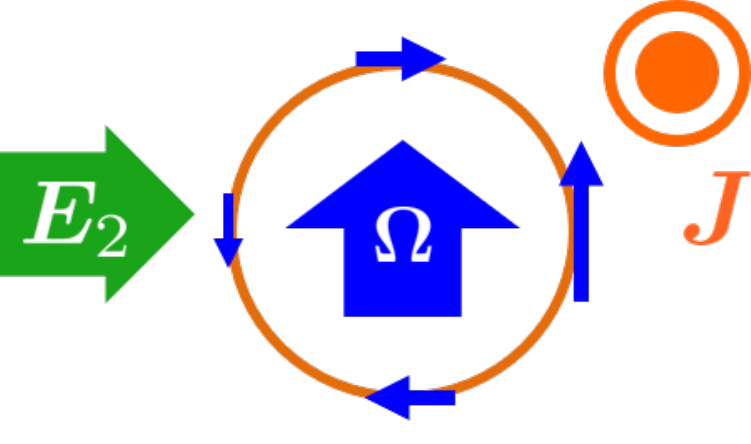} \\
			(a) &
			(b) &
			(c) 
			\end{tabular}
			\caption{A schematic picture of the nonlinear Hall effect due to the BCD. Orange-colored circles and blue-colored arrows represent a Fermi surface and Berry curvature, respectively. (a) In equilibrium, the Berry curvature  on the Fermi surface is completely compensated. (b) When an electric field $\bm{E}_1 \parallel \hat{x}$ is applied, the Fermi surface is shifted to the field direction, and the net Berry curvature $\bm{\Omega} \parallel \hat{y}$ is induced as in the case of the Edelstein effect. (c) Responding to the second electric field $\bm{E}_2 \parallel \hat{x}$, the electric current flowing along the $x$-axis is bent towards the $z$-axis owing to the anomalous Hall effect by the induced Berry curvature $\Omega^y$. This is an intuitive explanation of the BCD term, $\sigma_\mathrm{BCD}^{z;xx}$.}
			\label{Fig_BCD_symmetry_classification}
		\end{figure*}

The BCD term of nonlinear Hall conductivity can be understood by considering the spin-momentum locking [Fig.~\ref{Fig_BCD_symmetry_classification}\,(a)]. In the presence of the spin-momentum locking in electronic states, an electric field causes a shift of Fermi surfaces and accordingly changes the momentum-resolved spin polarization near Fermi surfaces. As a result, a net spin polarization is induced in a steady state. This is called Edelstein effect~\cite{Edelstein1990} and has been intensively discussed in recent spintronics research~\cite{Chernyshov2009}. We can intuitively understand the nonlinear Hall response arising from the BCD in an analogous way, as discussed below and illustrated in Fig.~\ref{Fig_BCD_symmetry_classification}.

For instance, let us consider a system with BCD, $\mathcal{D}^{\,xy} = \mathcal{D}^{\,yx}$, which is investigated in Sec.~\ref{Sec_magneticASOC_BCD}. 
Berry curvature on a Fermi surface can be illustrated as in Fig.~\ref{Fig_BCD_symmetry_classification}\,(a), where 
the total Berry curvature of occupied states, $\bm{\Omega} = \int d\bm{k} \sum_a  f(\epsilon_{\bm{k}a}) \bm{\Omega}_a(\bm{k})$, is completely canceled in the momentum space. At the first step, an applied electric field $\bm{E}_1 \parallel \hat{x}$ induces a shift of the Fermi surface and a total Berry curvature along the $y$ axis, $\Omega^y$, consequently emerges in a steady state [Fig.~\ref{Fig_BCD_symmetry_classification}\,(b)]. At the second step, because of the electric field $\bm{E}_2 \parallel \hat{x}$, the flowing electric current is bent towards the $z$-direction as it is by the anomalous Hall effect due to the dynamically-induced Berry curvature $\Omega^y$ [Fig.~\ref{Fig_BCD_symmetry_classification}\,(c)]. As a consequence, the nonlinear Hall response denoted by the component $\sigma_\mathrm{BCD}^{z;xx}$ occurs. All the other nonlinear Hall response coefficients induced by the BCD, $\mathcal{D}^{\,xy} = \mathcal{D}^{\,yx}$, can be explained by a similar argument. The allowed components satisfy the relation,
		\begin{equation}
		\sigma_\text{BCD}^{z;xx} = -\sigma_\text{BCD}^{z;yy}= -2\sigma_\text{BCD}^{x;xz}= 2\sigma_\text{BCD}^{y;yz}. \label{App_BCD_NLH_xy}
		\end{equation}
Note that a similar argument can be found in Ref.~\cite{Toshio2020Hydrodynbamics}. From the above-mentioned argument, we find that the chiral BCD given by $\trace{\hat{\mathcal{D}}}$ does not contribute to the nonlinear Hall response. In a system with a chiral BCD, $\mathcal{D}^{\,xx} = \mathcal{D}^{\,yy} = \mathcal{D}^{\,zz}$, the Berry curvature induced by an electric current is always parallel to the current. Then, the Hall response illustrated in Fig.~\ref{Fig_BCD_symmetry_classification} does not occur. The chiral BCD is, therefore, irrelevant to the nonlinear Hall response while the BCD itself can be finite in chiral crystals.

\subsection{Analytical and numerical results of BCD term}
In this section we calculate the BCD term of nonlinear Hall conductivity $\sigma_\text{BCD}$. 
First, we derive an analytical expression in simplified models. In the two-band Hamiltonian in Eq.~\eqref{two_band_Hamiltonian}, the Berry curvature is analytically obtained as
		\begin{equation}
		\Omega^\lambda_\pm = \mp \frac{1}{4} \epsilon_{\mu\nu\lambda} \left[  \partial_\mu \bm{\hat{g}} (\bm{k}) \times \partial_\nu \bm{\hat{g}} (\bm{k})  \right] \cdot \bm{\hat{g}} (\bm{k}),\label{2bands_Berry_curvature}
		\end{equation}
where the subscript $\pm$ represents the upper/lower band and $\bm{\hat{g}} (\bm{k})= \tilde{\bm{g}} (\bm{k})/|\tilde{\bm{g}}(\bm{k})|$ is an unit vector~\cite{Gao2014}. Compared to this case, the Berry curvature of the four band Hamiltonian [Eq.~\eqref{App_BMA_model_Hamiltonian}] for odd-parity magnetic multipole systems does not have a simple expression because of the inter-sublattice hopping term $V_\text{AB} (\bm{k})$. Therefore, we here adopt $V_\text{AB}(\bm{k}) = 0$ for simplicity to present analytical results. Although this simplification is not reasonable for \bma{}, it may be appropriate in other magnetic systems possessing valley or layer degree of freedom~\cite{Chittari2016}. Anyway an intuitive understanding for the BCD of magnetic multipole systems is obtained below.

Ignoring $V_\text{AB} (\bm{k})$, we obtain two pairs of bands labeled by the sublattice degree of freedom. The Berry curvature is given by
		\begin{equation}
		\Omega^\lambda_{(i)\pm} = \mp \frac{1}{4} \epsilon_{\mu\nu\lambda}  \left[  \partial_\mu \bm{\hat{g}}_{(i)} (\bm{k}) \times \partial_\nu \bm{\hat{g}}_{(i)} (\bm{k})   \right] \cdot \bm{\hat{g}}_{(i)} (\bm{k}),
		\end{equation}
where $i = (A,B)$ denotes the sublattice degree of freedom. The $g$-vector consists of the sASOC, the molecular field of antiferromagnetic order, and the Zeeman field,
		\begin{align}
		&\bm{g}_{(\text{A})} (\bm{k})= \bm{g}_0 (\bm{k}) + \bm{h}_\text{AF}+ \bm{h},\\
		&\bm{g}_{(\text{B})} (\bm{k})= -\bm{g}_0 (\bm{k}) - \bm{h}_\text{AF}+ \bm{h}.
		\end{align}
A symmetry analysis shows that the BCD, $\mathcal{D}^{\,xy} = \mathcal{D}^{\,yx}$, appears under the external magnetic field along the $z$-axis, $\bm{h}=h \hat{z}$. We therefore consider $x$ and $y$ components of the Berry curvature. For the model Hamiltonian in Eq.~\eqref{App_BMA_model_Hamiltonian} with the assumption that $V_\text{AB}(\bm{k}) = 0$ and $\alpha_2 =0$, the $x$-component is obtained as
		\begin{align}
		&\Omega^x_{(A)-}\notag \\ 
		&= +\frac{1}{2}  \left[ \partial_y \bm{\hat{g}}_{(\text{A})}  (\bm{k}) \times \partial_z \bm{\hat{g}}_{(\text{A})}  (\bm{k})  \right] \cdot \bm{\hat{g}}_{(\text{A})}   (\bm{k}),\\
		&= -\frac{1}{2|\bm{g}_{(\text{A})}  (\bm{k})|^3} \frac{\alpha_1^2\alpha_3}{2}\sin{k_x}\sin{\frac{k_x}{2}}\cos{k_y}\sin{\frac{k_y}{2}}\cos{\frac{k_z}{2}}.
		\end{align}
The model parameters in Eqs.~\eqref{bamn2as2modelparameter} and~\eqref{bamn2as2modelparameter2} imply 
		\begin{equation}
		|h_\text{AF}| \gg |t_i|,|\alpha_i|, |\bm{h}|,
		\end{equation}
and hence we perturbatively deal with the external magnetic field as
		\begin{equation}
		\frac{1}{|\bm{g}_{(\text{A})} (\bm{k})|^3} \simeq \frac{1}{|h_\text{AF}|^3} \left( 1 -\frac{3 \left[ \bm{g}_0 (\bm{k}) + \bm{h}  \right]\cdot \bm{h}_\text{AF}}{h_\text{AF}^2}  \right).
		\end{equation}
We neglect the term $\bm{g}_0 \cdot \bm{h}_\text{AF} / h_\text{AF}^{-3}$ at the right hand side since the term gives a small correction to the leading term. In the low density region, we have
		\begin{equation}
		\Omega^x_{(\text{A})-} \simeq - \frac{\alpha_1^2\alpha_3}{16| h_\text{AF}|^3} \left( 1- \frac{3h}{h_\text{AF}} \right) k_x^2 k_y.
		\end{equation}
Calculating $\Omega^x_{(\text{B})-}$ in the same way, we express the BCD summed over lower bands by
		\begin{align}
		&\mathcal{D}^{\,yx}_{(A)-} + \mathcal{D}^{\,yx}_{(B)-} \notag \\
			&= \int \frac{d\bm{k}}{\left( 2\pi \right)^d} \left[  \partial_y  \Omega^x_{(\text{A})-} f (\epsilon_{(\text{A})-}) + \partial_y  \Omega^x_{(\text{B})-} f (\epsilon_{(\text{B})-})  \right]\\
			&= -\frac{\alpha_1^2\alpha_3}{16 |h_\text{AF}|^3}  \int \frac{d\bm{k}}{\left( 2\pi \right)^d} k_x^2  \Biggl[  \left\{ f (\epsilon_{(\text{A})-}) -f(\epsilon_{(\text{B})-})  \right\} \notag \\
			&~~~~~-\frac{3h}{h_\text{AF}}\left\{ f (\epsilon_{(\text{A})-}) + f(\epsilon_{(\text{B})-})  \right\}  \Biggr],
		\end{align}
where the momentum dependence of energy $\epsilon_{(i)-}~(i = \text{A,~B})$ is implicit. The impact of the magnetic field is two-fold. One is the Fermi surface term [$f (\epsilon_{(\text{A})-}) -f(\epsilon_{(\text{B})-})$] determined by the Zeeman energy, and the other is the Fermi sea term [$f (\epsilon_{(\text{A})-}) + f(\epsilon_{(\text{B})-})$] derived from a correction to the Berry curvature proportional to the external field $h$. The former contribution is evaluated by
		\begin{align}
		&\int \frac{d\bm{k}}{\left( 2\pi \right)^d} k_x^2 \left\{   f(\epsilon_{(\text{A})-}) -f(\epsilon_{(\text{B})-}) \right\}\notag \\
		&\simeq \int \frac{d\bm{k}}{\left( 2\pi \right)^d} k_x^2 \frac{\partial f(\epsilon_{(\text{A})-})}{\partial \epsilon}\Bigr|_{h=0} (-2h), \\
		&=  \frac{mh}{6\pi^2}(2m|\epsilon_\text{F}|)^{3/2},
		\end{align}
where the energy spectrum at $h=0$ is approximated by the parabolic band dispersion, $\epsilon_{(\text{A})-} = -\bm{k}^2/2m$ in the last line, and $\epsilon_\text{F}$ is the Fermi energy of hole carriers. The Fermi sea term is evaluated as
		\begin{align}
		&-\frac{3h}{h_\text{AF}}\int \frac{d\bm{k}}{\left( 2\pi \right)^d} k_x^2 \left(   f(\epsilon_{(\text{A})-})+ f(\epsilon_{(\text{B})-}) \right) \notag \\
		&\simeq -\frac{6h}{h_\text{AF}} \int \frac{d\bm{k}}{\left( 2\pi \right)^d} k_x^2  [ f(\epsilon_{(\text{A})-}) ]\bigr|_{h=0}, \\
		&=  \frac{2mh}{5\pi^2}(2m|\epsilon_\text{F}|)^{3/2} \frac{|\epsilon_\text{F}|}{h_\text{AF}}.
		\end{align}
When the molecular field $h_\text{AF}$ is much larger than the Fermi energy $\epsilon_\text{F}$, the Fermi sea term is negligible compared to the Fermi surface term. Taking only the Fermi surface term, we obtain the magnetic-field-induced BCD,
		\begin{align}
		\mathcal{D}^{\,yx}_{(A)-} + \mathcal{D}^{\,yx}_{(B)-} 
			&\simeq  -\frac{\alpha_1^2\alpha_3}{16|h_\text{AF}|^3}  \frac{mh}{6\pi^2}(2m|\epsilon_\text{F}|)^{3/2}.\label{App_microscopic_calc_BCD_yx}
		\end{align}
From this expression the nonlinear Hall conductivity $\sigma_\text{BCD}^{z;xx}$ is given by
		\begin{equation}
		\sigma^{z;xx}_\text{BCD} = -\frac{q^3}{\gamma} \left( \mathcal{D}^{\,yx}_{(A)-} + \mathcal{D}^{\,yx}_{(B)-}  \right).
		\end{equation}
The magnetic-field-induced nonlinear Hall response is proportional to the external field and vanishes at $h=0$ where the \PT{}-symmetry is preserved. In other words, this nonlinear response is tunable by using the magnetic field. 

The presence of the magnetic-field-induced BCD and nonlinear Hall response is supported by numerical calculations. In the calculations, we take into account the inter-sublattice hopping term $V_\text{AB} (\bm{k})$ and assume the parameters in Eqs.~\eqref{bamn2as2modelparameter} and \eqref{bamn2as2modelparameter2}, and $h=0.01$. We confirmed that the numerical result is consistent with the symmetry analysis in Eq.~\eqref{App_BCD_NLH_xy}. Note that the nonlinear Hall conductivity is comparable between the cases $V_\text{AB} (\bm{k}) =0$ and $V_\text{AB} (\bm{k}) \neq 0$. Thus, above discussions for a simplified model are qualitatively appropriate.

\begin{figure}[htbp]
		\centering 
		\includegraphics[width=80mm,clip]{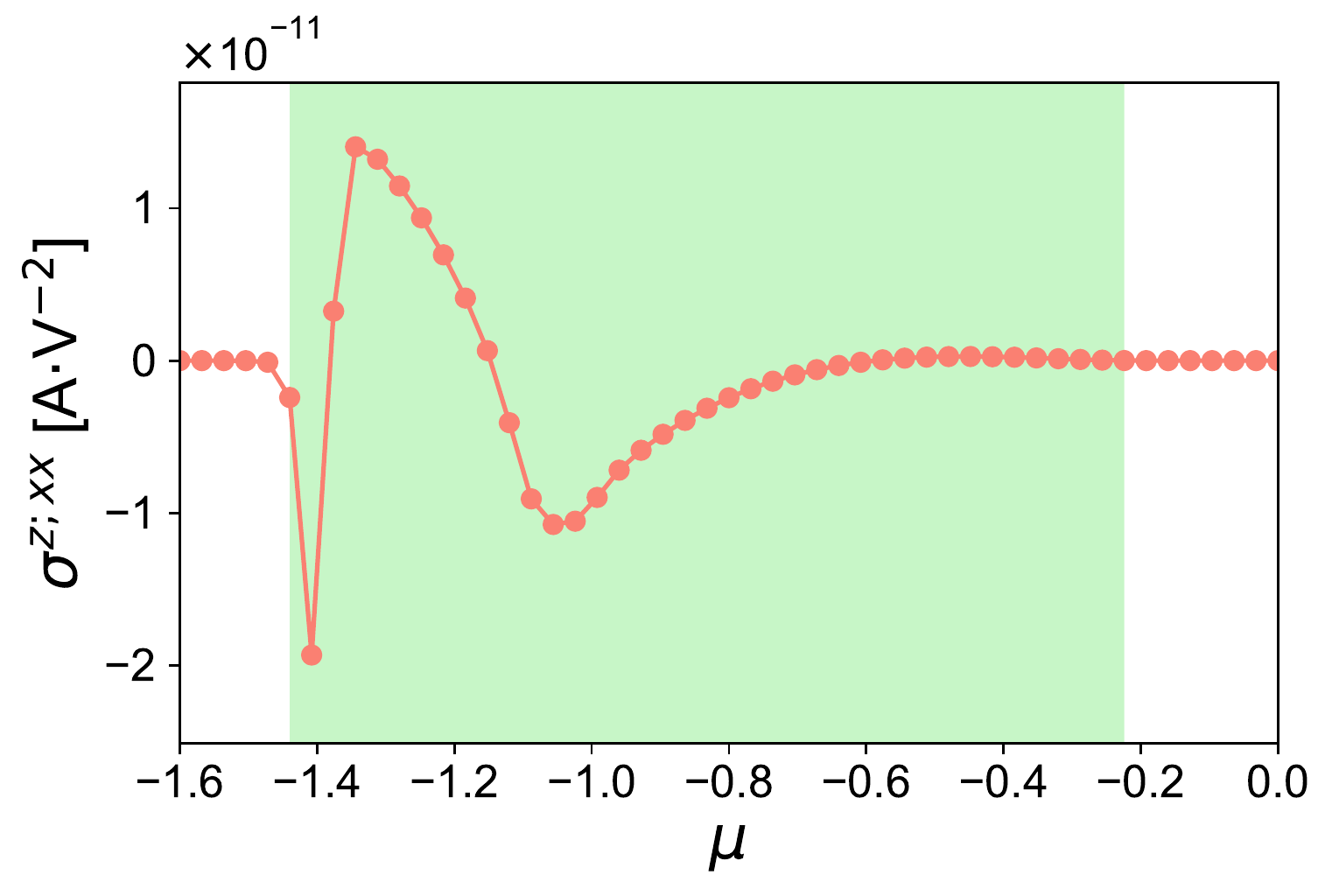}
		\caption{BCD term of nonlinear Hall conductivity $\sigma^{z;xx}_\text{BCD}$ under the magnetic field along the {\it z}-axis. Green-colored shaded region indicates the metallic region. The parameters are $T=0.01$, $\gamma^{-1}=1.0\times10^3$, and $h=0.01$. We take $N = 135^3$.}
		\label{Fig_BCD311_with_magnetic_field}
		\end{figure}

In Fig.~\ref{Fig_BCD311_with_magnetic_field} we show the numerical result of $\sigma_\text{BCD}^{z;xx}$  as a function of the chemical potential ranging around the lower energy bands [Eq.~\eqref{energyspectrum_with_magnetic_field}]. Because of the same reason as that for the nematicity-assisted dichroism shown in Fig.~\ref{Fig_Drude333_with_magnetic_field}, the nonlinear conductivity is small in the lightly hole-doped region. The magnitude is furthermore suppressed by the factor $\alpha_1^2\alpha_3/ |h_\text{AF}|^3$ revealed in Eq.~\eqref{App_microscopic_calc_BCD_yx}. A typical value is estimated as $\sigma^{z;xx}_\text{D} \sim  10^{-12}$ [A$\cdot$ V$^{-2}$] in our unit. The response may be enhanced when the exchange splitting is smaller than sASOC or comparable to it. Such situation realizes in antiferromagnetic materials including \bma{} near N\'{e}el temperatures.

\section{Extrinsic contribution to nonlinear conductivity}\label{App_Sec_extrinsic}

The formulation presented in Sec.~\ref{App_Sec_derivation_NLC} is based on the clean limit ($\tau \rightarrow \infty$) with the phenomenological scattering rate, and thus intrinsic contributions are studied throughout our work. On the other hand, extrinsic contributions such as impurity scattering and electron correlations may be non-negligible in a realistic situation. For instance, it is well-known that the anomalous Hall effect is affected by impurities and that such extrinsic contributions may overwhelm the intrinsic contributions~\cite{Tian2009,Nagaosa2010}. Moreover, a recent experiment pointed out that the impurity scattering contributes to the nonlinear Hall response in WTe$_2$ as well as the Berry curvature dipole~\cite{Kang2019}. Therefore, it is necessary to examine whether our classification shown in Table~\ref{Table_relaxation_time_dependence_2nd_conductivity} is modified when extrinsic contributions included.

According to a recent theoretical work~\cite{Du2019}, the formula for NLC is modified by the skew scattering and side jump effects arising from the impurity scattering. Ref.~\cite{Du2019} clearly shows that the NLC in \T{}-symmetric systems includes such extrinsic terms in addition to the Berry curvature dipole effect in Eq.~\eqref{nonlinear_conductivity_Berry_curvature_dipole_term} and that the Drude term appears as a result of the \T{}-symmetry breaking. On the other hand, it has not been studied whether the nonlinear conductivity formula in \PT{}-symmetric systems is modified in the presence of the impurity scattering. Thus, we below consider the effect of impurity scattering in the \PT{}-preserved systems. Interestingly, we will see that the \PT{}-symmetry leads to strong suppression of the extrinsic contributions. 

For simplicity, we consider spinless systems and take into account on-site impurities having the delta-function-shaped potential. The potential energy of the randomly-distributed impurities is written by
		\begin{equation}
		V_\text{imp} ( \hat{\bm{r}}) = \sum_i v_0 \delta (\hat{\bm{r}} -\bm{R}_i), \label{app_impurity_potential}
		\end{equation}
where $v_0$ denotes the strength of impurity potential, $i$ labels the impurities, and $\bm{R}_i$ denotes the position of the $i$-th impurity. Owing to the random distribution of impurities, the random average $\impave{}$ satisfies the relation
		\begin{equation}
		\impave{V_\text{imp}} = 0.
		\end{equation}
We below omit the subscript `imp' of $V_\text{imp}$ unless otherwise mentioned. Based on the results in Ref.~\cite{Du2019}, the leading extrinsic contributions to the NLC $\sigma^{\mu;\nu\lambda}_\text{ext}$ are three-fold:
		\begin{equation}
		\sigma^{\mu;\nu\lambda}_\text{ext} = \sigma_\text{sj,1}^{\mu;\nu\lambda}+ \sigma_\text{sj,2}^{\mu;\nu\lambda}+ \sigma_\text{sk}^{\mu;\nu\lambda},\label{App_extrinsic_NLC}
		\end{equation}
in which the abbreviated labels `sj` and 'sk' represent `side jump' and `skew scattering' effects~\cite{Du2019}, respectively. These contributions are added to the Berry curvature dipole term in the \T{}-symmetric systems and may explain the observed nonlinear Hall response in WTe$_2$~\cite{Kang2019}.

First, we investigate the side jump contributions. The side jump terms are determined by the side jump velocity given by
		\begin{equation}
		v^\mu_{\text{(sj)}\bm{a}} = \sum_{\bm{b}} \mathcal{W}^\text{sy}_{\bm{a}\bm{b}} \delta r^\mu_{\bm{b}\bm{a}},
		\end{equation}
where $\bm{a} = (a,\bk_a)$ [$\bm{b} = (b,\bk_b )$] denotes Bloch states labeled by the band index $a$ ($b$) and crystal momentum $\bk_a$ ($\bk_b$). $\mathcal{W}^\text{sy}_{\bm{a}\bm{b}} = \mathcal{W}_{\bm{a}\bm{b}}/2 +\mathcal{W}_{\bm{b}\bm{a}}/2  $ represents the symmetric part of the scattering amplitude $\mathcal{W}_{\bm{a}\bm{b}}$. $\mathcal{W}_{\bm{a}\bm{b}}$ is defined with the T-matrix as 
		\begin{equation}
		\mathcal{W}_{\bm{a}\bm{b}} = \frac{2\pi}{\hbar} |T_{\bm{a}\bm{b}}|^2 \delta \left( \epsilon_{\bm{a}\bm{b}} \right),
		\end{equation} 
in which we introduced $\epsilon_{\bm{a}\bm{b}} = \epsilon_{\bk_a a} - \epsilon_{\bk_b b}$. Importantly, the side jump velocity depends on the positional shift $\delta r^\mu_{\bm{a}\bm{b}}$ given by
		\begin{align}
		\delta r^\mu_{\bm{a}\bm{b}} 
			&= \Braket{u_a (\bk_a) | i \frac{\partial u_a}{\partial k_a^\mu} (\bk_a)} - \Braket{u_b  (\bk_b)| i \frac{\partial u_b}{\partial {k}_b^\mu}(\bk_b) } \notag \\
			&~~ - \left( \frac{\partial}{\partial k_a^\mu} + \frac{\partial}{\partial k_b^\mu}  \right) \text{arg}\left( V_{\bm{a}\bm{b}} \right),
		\end{align}
which represents the coordinate shift during the scattering process $\bm{a} \leftarrow \bm{b}$~\cite{Sinitsyn2006}. With the impurity potential in Eq.~\eqref{app_impurity_potential}, the matrix element of $V$ is given by
		\begin{align}
		V_{\bm{a}\bm{b}}
			&= c v_0 \Braket{u_{a} (\bk_a) | u_{b} (\bk_b)} \sum_i e^{-i \left( \bk_a-\bk_b \right)\cdot \bm{R}_i },\\
			&= c v_0 I_{\bm{a}\bm{b}} \rho_{\bk_a-\bk_b},
		\end{align}
where $c$ denotes a scalar constant, $I_{\bm{a}\bm{b}} = \Braket{u_a (\bk_a) | u_b (\bk_b)}$, and $\rho_{\bm{Q}} = \sum_i \exp{\left( -i \bm{Q}\cdot \bm{R}_i \right)} $. Since the matrix element $V_{\bm{a}\bm{b}}$ depends on $\bk_a -\bk_b$, the positional shift is simplified as
		\begin{align}
		\delta r^\mu_{\bm{a}\bm{b}} 
			&\rightarrow   \Braket{u_a (\bk_a) | i \frac{\partial u_a}{\partial k_a^\mu} (\bk_a)} - \Braket{u_b  (\bk_b)| i \frac{\partial u_b}{\partial {k}_b^\mu}(\bk_b) }\notag \\
			&~~- \left( \frac{\partial}{\partial k_a^\mu} + \frac{\partial}{\partial k_b^\mu}  \right) \text{arg}\,\Braket{u_{a} (\bk_a) | u_{b} (\bk_b)},
		\end{align}
which is similar to the so-called shift vector~\cite{Sturman1992Book}. Now, we consider the constraint due to the \PT{}-symmetry. The Bloch state is transformed by \PT{}-symmetry into
		\begin{equation}
		\ket{u_a (\bk_a)} \rightarrow \overline{\ket{u_a (\bk_a)}} = \ket{u_a (\bk_a)} e^{-i\phi_a (\bk_a)},
		\end{equation}
where $\phi_a$ denotes the phase factor determined by the adopted gauge. Accordingly, we obtain the relation of the connection term $\Braket{u_a (\bk_a)| i \partial_\mu u_a(\bk_a) }$ given by
		\begin{align}
		&\Braket{u_a (\bk_a)| i \frac{\partial u_a}{\partial k_a^\mu}(\bk_a)}\notag \\
		&= i \overline{\Braket{\frac{\partial u_a}{\partial k_a^\mu}(\bk_a)  |u_a (\bk_a) }},\\
		&= -\Braket{u_a (\bk_a)| i \frac{\partial u_a}{\partial k_a^\mu} (\bk_a)} - \frac{\partial \phi_a (\bk_a)}{\partial k_a^\mu}.
		\end{align}
Similarly, using the \PT{}-symmetry, we can see 
		\begin{align}
		&\text{arg}\,\Braket{u_{a} (\bk_a) | u_{b} (\bk_b)} \notag \\
		&= - \text{arg}\, \Braket{u_a (\bk_a)| u_{b} (\bk_b)} + \phi_b (\bk_b) - \phi_a (\bk_a).
		\end{align}
Thus, the positional shift satisfies
		\begin{equation}
		\delta r^\mu_{\bm{a}\bm{b}} = -\delta r^\mu_{\bm{a}\bm{b}} = 0,
		\end{equation}
which indicates that the side jump velocity $v^\mu_\text{(sj)}$ vanishes by the \PT{}-symmetry. As a result, the side jump contributions $ \sigma_\text{sj,1}$ and $\sigma_\text{sj,2}$ in Eq.~\eqref{App_extrinsic_NLC} are forbidden in the \PT{}-symmetric systems. We can also understand this conclusion intuitively from the fact that the positional shift $\delta r^\mu_{\bm{a}\bm{b}}$ occurring in the forward scattering roughly corresponds to $\sim \epsilon_{\mu \nu\lambda} \Omega_\nu (k_a^\lambda -k_b^\lambda)$ where we introduce the Berry curvature $\bm{\Omega}$~\cite{Sinitsyn2006}. We can see immediately that $\delta r^\mu_{\bm{a}\bm{b}}= 0 $ since $\bm{\Omega}=0$ due to the \PT{}-symmetry. Although the derivation has assumed the delta-function-shaped impurity potential, the positional shift is generally suppressed by the \PT{}-symmetry in the case of the spherically-symmetric impurity potential whose Fourier component $V_{\bm{a}\bm{b}}$ is a function of $\bk_a-\bk_b$. 

Next, we consider the skew scattering contribution denoted by $\sigma_\text{sk}$ in Eq.~\eqref{App_extrinsic_NLC}. Although we do not show the expression of the skew scattering contribution (Ref.~\cite{Du2019} for details), an important ingredient taking the whole expression is the anti-symmetric component of the scattering amplitude $\mathcal{W}_{\bm{a}\bm{b}}^\text{as} = \mathcal{W}_{\bm{a}\bm{b}}/2 -\mathcal{W}_{\bm{b}\bm{a}}/2$. Thus, the skew scattering contribution is closely related to the anti-symmetric scattering between the states $\bm{a}$ and $\bm{b}$. 

To evaluate $\mathcal{W}^\text{as}$, perturbation expansion of the T-matrix is performed here. The T-matrix is defined as
		\begin{equation}
		T_{\bm{a}\bm{b}} = \Braket{\psi_{\bk_a a} | \hat{V} | \Psi_{\bm{b}}},
		\end{equation} 
where the ket vector $\ket{\Psi_{\bm{b}}}$ is obtained by solving Lipman-Schwinger equation with the impurity potential $\hat{V}$. The equation is written as
		\begin{equation}
		\ket{\Psi_{\bm{b}}} = \ket{\psi_{\bk_b b}} + \frac{\hat{V}}{\epsilon_{\bk_b b} - \hat{H}_0 + i\eta} \ket{\Psi_{\bm{b}}},
		\end{equation}
where $\hat{H}_0$ denotes the unperturbed Hamiltonian and $\eta\, (>0)$ is an infinitesimal number. We therefore obtain perturbation expansion of the T-matrix in the power of the matrix element $V_{\bm{a}\bm{b}}$. The lowest-order contribution is obtained by replacing $T_{\bm{a}\bm{b}}$ with $V_{\bm{a}\bm{b}}$, that is, $\ket{\Psi_{\bm{b}}}\rightarrow \ket{\psi_{\bk_b b}}$. The resulting contribution to $\mathcal{W}_{\bm{a}\bm{b}}$ is given by
		\begin{equation}
 		\mathcal{W}^{(2)}_{\bm{a}\bm{b}} = \frac{2\pi}{\hbar} |V_{\bm{a}\bm{b}}|^2 \delta \left( \epsilon_{\bm{a}\bm{b}} \right),
 		\end{equation} 
which is $O(V^2)$. The expression is symmetric under the permutation $\bm{a}\leftrightarrow \bm{b}$ and therefore this term does not yield a skew scattering. The third-order contribution to $\mathcal{W}_{\bm{a}\bm{b}}$ is given by
		\begin{align}
		&\frac{\hbar}{2\pi} \mathcal{W}^{(3)}_{\bm{a}\bm{b}} \notag \\
			&=  \sum_{\bm{c}}\left( \frac{\Braket{V_{\bm{a}\bm{b}}^\ast V_{\bm{a}\bm{c}}V_{\bm{c}\bm{b}}}_\text{imp}}{\epsilon_{\bm{b}\bm{c}} +i\eta } + \frac{\Braket{V_{\bm{a}\bm{c}}^\ast V_{\bm{c}\bm{b}}^\ast V_{\bm{a}\bm{b}}}_\text{imp}}{\epsilon_{\bm{a}\bm{c}} - i\eta }  +c.c \right) \delta \left( \epsilon_{\bm{a}\bm{b}} \right),\\
			&=  \sum_{\bm{c}}\left( \frac{\Braket{V_{\bm{a}\bm{b}}^\ast V_{\bm{a}\bm{c}}V_{\bm{c}\bm{b}}}_\text{imp}}{\epsilon_{\bm{b}\bm{c}} +i\eta } + \frac{\Braket{V_{\bm{a}\bm{c}}^\ast V_{\bm{c}\bm{b}}^\ast V_{\bm{a}\bm{b}}}_\text{imp}}{\epsilon_{\bm{b}\bm{c}} - i\eta }  +c.c \right) \delta \left( \epsilon_{\bm{a}\bm{b}} \right). \label{app_scattering_amplitude_3rd}
		\end{align}
Considering the impurity potential defined in Eq.~\eqref{app_impurity_potential}, we have
		\begin{equation}
		\Braket{V_{\bm{a}\bm{b}}^\ast V_{\bm{a}\bm{c}}V_{\bm{c}\bm{b}}}_\text{imp} \propto I_{\bm{b}\bm{a}} I_{\bm{a}\bm{c}} I_{\bm{c}\bm{b}}  \impave{\rho_{\bk_a-\bk_b}^\ast\rho_{\bk_a-\bk_c}\rho_{\bk_c-\bk_b}}.
		\end{equation}
By taking the average over random distribution of impurities~\cite{KohnLuttinger1957Transport}, we obtain
		\begin{align}
		&\impave{\rho_{\bk_1}\rho_{\bk_2}\rho_{\bk_2}} \notag \\
			&=N_\text{imp}\delta_{\bk_1 + \bk_2+ \bk_3,0}  \notag \\
			&+N_\text{imp}\left( N_\text{imp}-1 \right)\notag \\
			&~~\times \left( \delta_{\bk_1,0}\delta_{\bk_2 + \bk_3,0}+ \delta_{\bk_2,0}\delta_{\bk_3 + \bk_1,0} + \delta_{\bk_3,0}\delta_{\bk_1 + \bk_2,0} \right)\notag \\
			& + N_\text{imp}\left( N_\text{imp}-1 \right)\left( N_\text{imp}-2 \right) \delta_{\bk_1,0}\delta_{\bk_2,0}\delta_{\bk_3,0},
		\end{align}
where $N_\text{imp}$ denotes the total number of impurities. Thus, the component in $\impave{}$ of Eq.~\eqref{app_scattering_amplitude_3rd} is symmetric under $\bk_a \leftrightarrow \bk_b$. We also obtain the relation
		\begin{equation}
		I_{\bm{b}\bm{a}} I_{\bm{a}\bm{c}} I_{\bm{c}\bm{b}} = I_{\bm{a}\bm{b}} I_{\bm{c}\bm{a}} I_{\bm{b}\bm{c}},
		\end{equation}
by making use of the \PT{}-symmetry. It is therefore shown that the first term in Eq.~\eqref{app_scattering_amplitude_3rd} is symmetric under $\bm{a} \leftrightarrow	\bm{b}$ and does not contribute to $\mathcal{W}^{\text{(as)}}$. Performing the parallel discussion for the remaining terms in Eq.~\eqref{app_scattering_amplitude_3rd}, we can see $\mathcal{W}^{(3,\text{as})}_{\bm{a}\bm{b}} \equiv \mathcal{W}^{(3)}_{\bm{a}\bm{b}}/2- \mathcal{W}^{(3)}_{\bm{b}\bm{a}}/2=0$, indicating the absence of the third-order contribution to the skew scattering term. In a similar manner, we can prove that the fourth-order term $\mathcal{W}^{(4)}_{\bm{a}\bm{b}}$ does not have its anti-symmetric component. 
Although we do not calculate higher-order terms, the cancellation may also happen in higher-order terms in $V_\text{imp}$. 
From the above discussions, we conclude that the anti-symmetric scattering $\mathcal{W}^{\text{(as)}}$ and the resulting skew scattering term are strongly suppressed by the \PT{}-symmetry. 

To summarize this section, extrinsic contributions to second-order NLC are strongly suppressed by the \PT{}-symmetry, while they play an important role in the \T{}-symmetric systems~\cite{Du2019,Kang2019}. We therefore expect that the symmetry classification of nonlinear conductivity (Table~\ref{Table_relaxation_time_dependence_2nd_conductivity}) remains meaningful beyond the relaxation time approximation when we focus on the \PT{}-symmetric magnetic systems.      

\bibliography{nonlinear_conductivity}

\end{document}